\pdfoutput=1

\documentclass[12pt,a4paper]{article}

\usepackage{ifthen}
\newboolean{pdflatex}
\setboolean{pdflatex}{true}

\newboolean{articletitles}
\setboolean{articletitles}{true}

\newboolean{uprightparticles}
\setboolean{uprightparticles}{false}

\def\paperauthors{LHCb collaboration}
\def\paperasciititle{Production of eta and eta' mesons in pp and pPb collisions}
\def\papertitle{Production of $\eta$ and $\eta'$ mesons in $pp$ and $p$Pb collisions}
\def\paperkeywords{{High Energy Physics}, {LHCb}}
\def\papercopyright{\the\year\ CERN for the benefit of the LHCb collaboration}
\def\paperlicence{CC BY 4.0 licence}
\def\paperlicenceurl{https://creativecommons.org/licenses/by/4.0/}

\usepackage[top=1in, bottom=1.25in, left=1in, right=1in]{geometry}

\usepackage{microtype}
\usepackage{lineno}
\usepackage{xspace}
\usepackage{caption}

\usepackage{graphicx}
\usepackage{color}
\usepackage{colortbl}

\usepackage{amsmath}
\usepackage{amssymb}
\usepackage{amsfonts}
\usepackage{upgreek}

\newcommand*\patchAmsMathEnvironmentForLineno[1]{%
\expandafter\let\csname old#1\expandafter\endcsname\csname #1\endcsname
\expandafter\let\csname oldend#1\expandafter\endcsname\csname
end#1\endcsname
 \renewenvironment{#1}%
   {\linenomath\csname old#1\endcsname}%
   {\csname oldend#1\endcsname\endlinenomath}%
}
\newcommand*\patchBothAmsMathEnvironmentsForLineno[1]{%
  \patchAmsMathEnvironmentForLineno{#1}%
  \patchAmsMathEnvironmentForLineno{#1*}%
}
\AtBeginDocument{%
\patchBothAmsMathEnvironmentsForLineno{equation}%
\patchBothAmsMathEnvironmentsForLineno{align}%
\patchBothAmsMathEnvironmentsForLineno{flalign}%
\patchBothAmsMathEnvironmentsForLineno{alignat}%
\patchBothAmsMathEnvironmentsForLineno{gather}%
\patchBothAmsMathEnvironmentsForLineno{multline}%
\patchBothAmsMathEnvironmentsForLineno{eqnarray}%
}

\usepackage{hyperxmp}

\usepackage[pdftex,
            pdfauthor={\paperauthors},
            pdftitle={\paperasciititle},
            pdfkeywords={\paperkeywords},
            pdfcopyright={Copyright (C) \papercopyright},
            pdflicenseurl={\paperlicenceurl}]{hyperref}

\usepackage[bottom,flushmargin,hang,multiple]{footmisc}

\usepackage[all]{hypcap}

\def\lhcb   {\mbox{LHCb}\xspace}
\newcommand{\aunit}[1]{\ensuremath{\text{\,#1}}} 
\newcommand{\tev}{\aunit{Te\kern -0.1em V}\xspace}
\newcommand{\gev}{\aunit{Ge\kern -0.1em V}\xspace}
\newcommand{\mev}{\aunit{Me\kern -0.1em V}\xspace}
\def\PK      {\ensuremath{K}\xspace}
\def\kaon    {{\ensuremath{\PK}}\xspace}
\def\KS      {{\ensuremath{\kaon^0_{\mathrm{S}}}}\xspace}
\def\pythia     {\mbox{\textsc{Pythia}}\xspace}
\def\photos     {\mbox{\textsc{Photos}}\xspace}
\def\geant      {\mbox{\textsc{Geant4}}\xspace}
\def\evtgen     {\mbox{\textsc{EvtGen}}\xspace}

\newcommand{\lhcborcid}[1]{\href{https://orcid.org/#1}{\hspace*{0.1em}\raisebox{-0.45ex}{\includegraphics[width=1em]{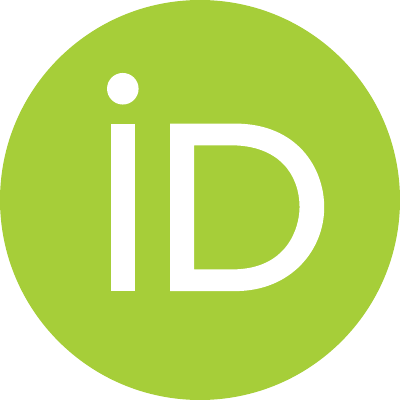}}}}

\usepackage{longtable}
\usepackage{cite}
\usepackage{mciteplus}

\begin{document}

\renewcommand{\thefootnote}{\fnsymbol{footnote}}
\setcounter{footnote}{1}

\begin{titlepage}
\pagenumbering{roman}

\vspace*{-1.5cm}
\centerline{\large EUROPEAN ORGANIZATION FOR NUCLEAR RESEARCH (CERN)}
\vspace*{1.5cm}
\noindent
\begin{tabular*}{\linewidth}{lc@{\extracolsep{\fill}}r@{\extracolsep{0pt}}}
\ifthenelse{\boolean{pdflatex}}
{\vspace*{-1.5cm}\mbox{\!\!\!\includegraphics[width=.14\textwidth]{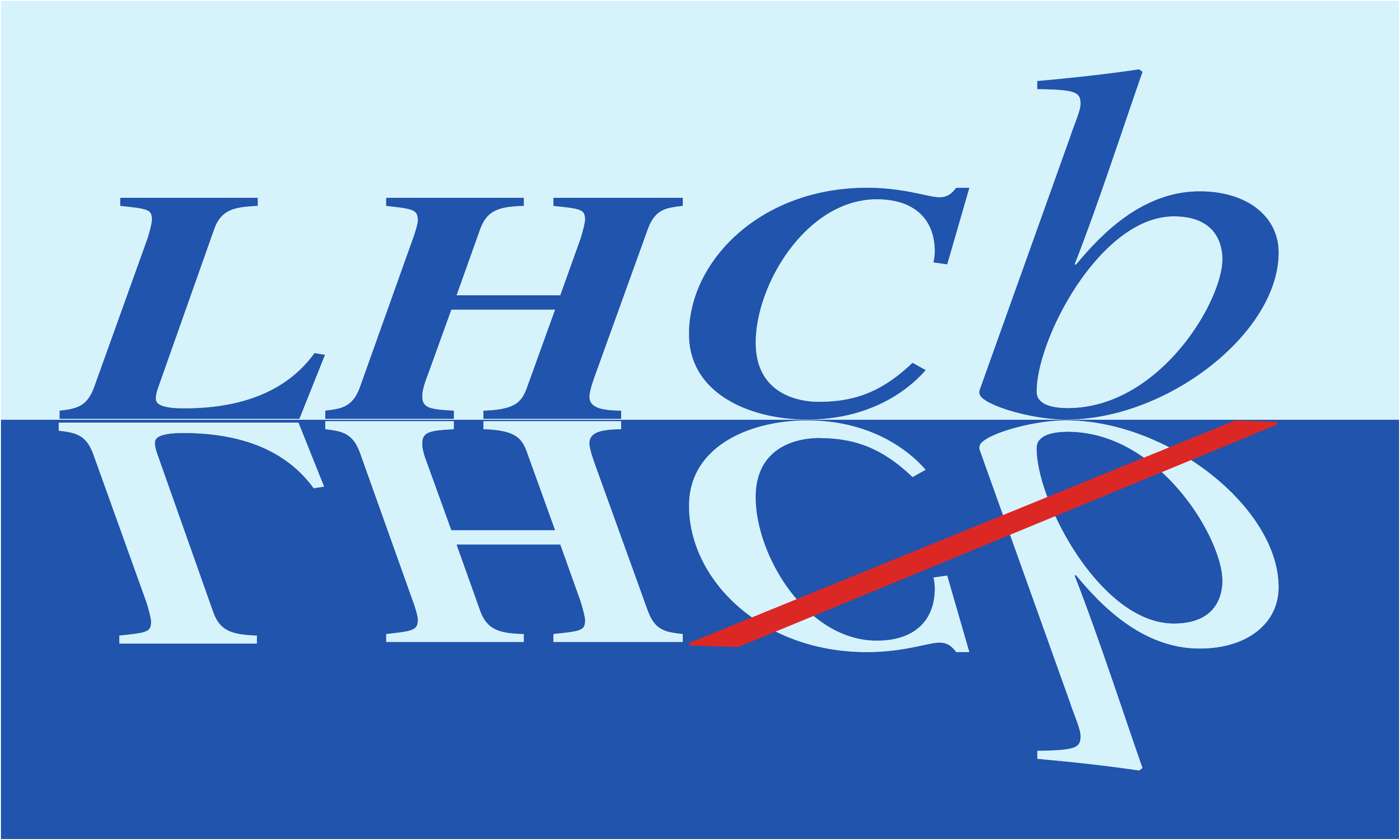}} & &}%
{\vspace*{-1.2cm}\mbox{\!\!\!\includegraphics[width=.12\textwidth]{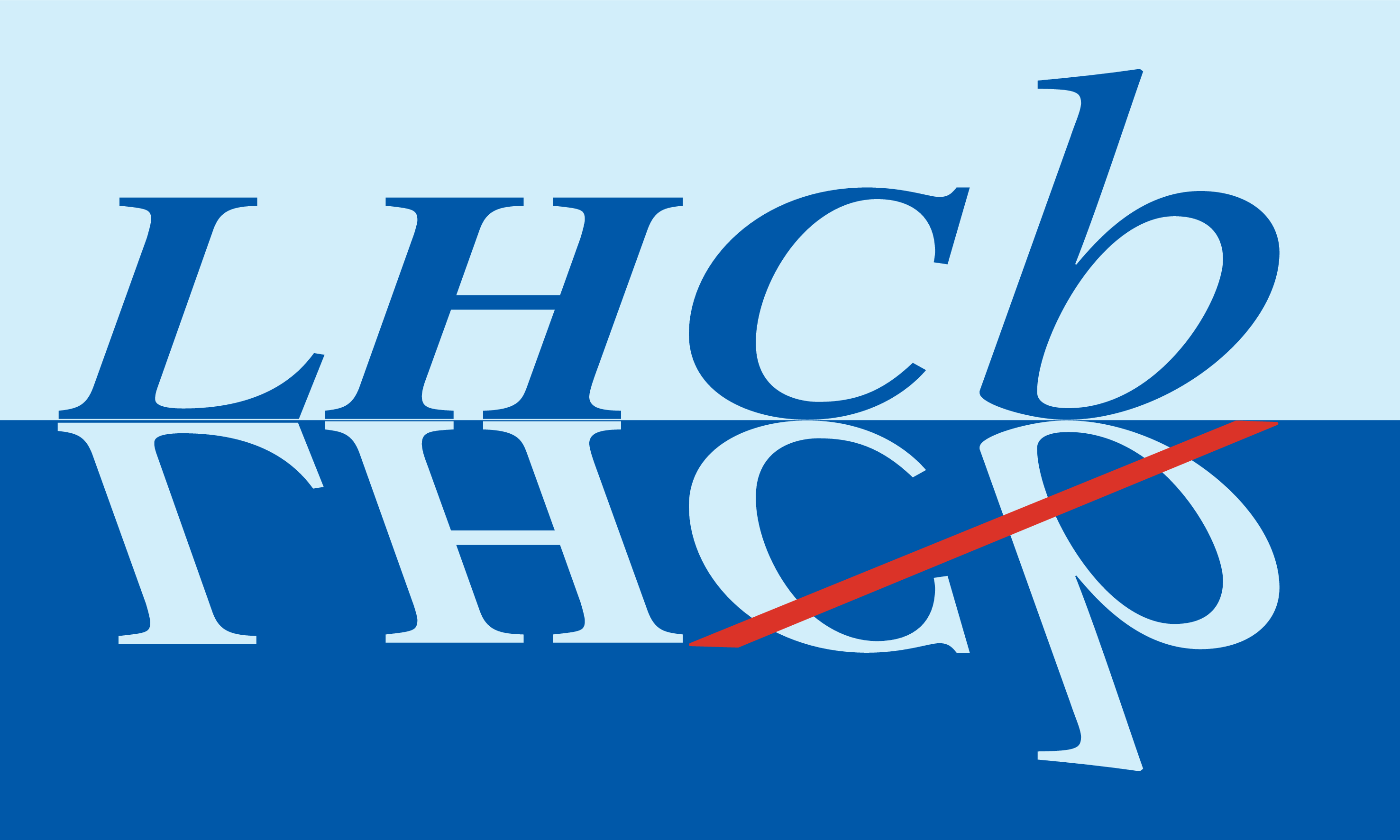}} & &}%
\\
 & & CERN-EP-2023-228 \\
 & & LHCb-PAPER-2023-030 \\
 & & February 13, 2025 \\
 & & \\
\end{tabular*}

\vspace*{3.0cm}

{\normalfont\bfseries\boldmath\huge
\begin{center}
  \papertitle 
\end{center}
}

\vspace*{2.0cm}

\begin{center}
\paperauthors\footnote{Authors are listed at the end of this paper.}
\end{center}

\vspace{\fill}

\begin{abstract}
  \noindent The production of $\eta$ and $\eta'$ mesons is studied in
  proton-proton and proton-lead collisions collected with the LHCb
  detector. Proton-proton collisions are studied at center-of-mass
  energies of $5.02$ and $13~{\rm TeV}$ and proton-lead collisions are
  studied at a center-of-mass energy per nucleon of $8.16~{\rm
  TeV}$. The studies are performed in center-of-mass (c.m.) rapidity
  regions \mbox{$2.5<y_{\rm c.m.}<3.5$} (forward rapidity)
  and \mbox{$-4.0<y_{\rm c.m.}<-3.0$} (backward rapidity) defined
  relative to the proton beam direction. The $\eta$ and $\eta'$
  production cross sections are measured differentially as a function
  of transverse momentum for \mbox{$1.5<p_{\rm T}<10~{\rm GeV}$}
  and \mbox{$3<p_{\rm T}<10~{\rm GeV}$}, respectively. The
  differential cross sections are used to calculate nuclear
  modification factors. The nuclear modification factors for $\eta$
  and $\eta'$ mesons agree at both forward and backward rapidity,
  showing no significant evidence of mass dependence. The differential
  cross sections of $\eta$ mesons are also used to calculate
  $\eta/\pi^0$ cross section ratios, which show evidence of a
  deviation from the world average. These studies offer new
  constraints on mass-dependent nuclear effects in heavy-ion
  collisions, as well as $\eta$ and $\eta'$ meson fragmentation.
\end{abstract}

\vspace*{2.0cm}

\begin{center}
  Published in Phys. Rev. C109 (2024), 024907
\end{center}

\vspace{\fill}

{\footnotesize 
\centerline{\copyright~\papercopyright. \href{\paperlicenceurl}{\paperlicence}.}}
\vspace*{2mm}

\end{titlepage}

\newpage
\setcounter{page}{2}
\mbox{~}

\renewcommand{\thefootnote}{\arabic{footnote}}
\setcounter{footnote}{0}

\cleardoublepage

\pagestyle{plain}
\setcounter{page}{1}
\pagenumbering{arabic}

\section{Introduction}

Light hadron production in heavy-ion collisions is a sensitive probe of the
structure of the colliding nuclei. Pion production data from deuteron-gold
($d$-Au) collisions at the Relativistic Heavy Ion Collider (RHIC) have been used
to constrain nuclear parton distribution functions
(nPDFs)~\cite{Eskola:2016oht,deFlorian:2011fp,PHENIX:2006mhb,STAR:2006xud,STAR:2009qzv}.
These nPDFs encode modifications of the partonic structure of nuclei within the
collinear factorization framework~\cite{Collins:1989gx}. In addition, light
hadron production is also sensitive to effects of the high energy densities
produced in heavy-ion collisions. Light hadron production was one of the
original tools used to study quark-gluon plasma (QGP) in heavy-ion collisions.
Measurements of angular correlations between light hadrons in heavy-ion
collisions at RHIC demonstrated collective flow and were interpreted as some of
the first evidence for QGP
production~\cite{BRAHMS:2004adc,PHOBOS:2004zne,STAR:2005gfr,PHENIX:2004vcz}.
Similar correlations have been observed in small-collision systems at the Large
Hadron Collider (LHC) and RHIC, pointing to possible QGP
formation~\cite{CMS:2010ifv,ALICE:2012eyl,ATLAS:2012cix,CMS:2012qk,LHCb-PAPER-2015-040,PHENIX:2013ktj,PHENIX:2018lia}.

Nuclear effects on particle production can be quantified using the nuclear
modification factor, which is defined as
\begin{equation}
    R_{p{\rm Pb}}=\frac{1}{A}\frac{d\sigma_{p{\rm Pb}}/dp_{\rm T}}{d\sigma_{pp}/dp_{\rm T}},
    \label{eqn:rpa}
\end{equation}
where $A=208$ is the atomic mass of the lead nucleus, while
$d\sigma_{p{\rm Pb}}/dp_{\rm T}$ and $d\sigma_{pp}/dp_{\rm T}$ are the
differential cross sections as a function of transverse momentum $p_{\rm T}$
in $p{\rm Pb}$ and $pp$ collisions, respectively. The LHCb experiment recently
measured the nuclear modification factors of inclusive charged particles in
proton-lead ($p{\rm Pb}$) collisions at $\sqrt{s_{\rm NN}}=5.02\,{\rm TeV}$ and
$\pi^0$ meson production at $8.16\,{\rm TeV}$, observing enhancements in the
lead-going direction larger than those predicted by nPDF
calculations~\cite{LHCb-PAPER-2021-015,LHCb-PAPER-2021-053}. These enhancements
could be produced by effects such as hydrodynamic evolution of QGP created in
these collisions~\cite{Pierog:2013ria}. Understanding the origin of QGP-like
effects in small collision systems is one of the primary goals of high-energy
nuclear physics.

Studying the production of $\pi^0$, $\eta$, and $\eta'$ mesons in heavy-ion
collisions allows for the isolation of the mass and isospin dependence of
nuclear effects, which can help reveal the origin of QGP-like phenomena in
small-collision systems. Collective radial flow of the QGP, for example, is
expected to produce larger enhancements for heavier particles, as heavier
particles must receive a larger momentum boost in order to comove with an
expanding medium~\cite{Sickles:2013yna,Ayala:2006bc}. Studies of the
multiplicity dependence of identified particle production in proton-proton
($pp$) collisions by the ALICE collaboration indicate a hardening of the $p_{\rm
T}$ spectra with increasing multiplicity that is more pronounced for baryons and
strange mesons than for pions~\cite{ALICE:2020nkc,ALICE:2018pal}. The authors
interpret these results as a mass ordering consistent with radial flow. This
interpretation is further complicated, however, by differences in the number and
flavor of valence quarks between the hadron species. The $\eta$ and $\eta'$
mesons have similar masses to the kaon and proton, respectively, but with
different valence quark content. As a result, studies of production of $\eta$
and $\eta'$ mesons can clarify the origin of these flow-like effects.

The production of $\eta$ mesons has been studied extensively in small-collision
systems at central rapidity at RHIC and the
LHC~\cite{PHENIX:2010qqf,PHENIX:2010hvs,ALICE:2012wos,ALICE:2017ryd,ALICE:2018vhm,ALICE:2021est,STAR:2009qzv}.
In addition, the STAR and PHENIX collaborations have studied $\eta$ meson
production at forward rapidity in polarized $pp$ collisions at
$\sqrt{s}=200\gev{}$~\cite{STAR:2012ljf,PHENIX:2014qwb}. There are no studies,
however, of $\eta$ meson production at forward or backward rapidity in
collisions involving heavy ions. Furthermore, $\eta'$ meson production has only
been studied in $pp$ collisions at $\sqrt{s}=200\gev{}$ by the PHENIX
collaboration~\cite{PHENIX:2010qqf}. The $\eta'$ nuclear modification factor has
never been measured in heavy-ion collisions. Studying both $\eta$ and $\eta'$
production at forward and backward rapidities helps reveal the mass and rapidity
dependence of nuclear effects in heavy-ion collisions.

This article presents measurements of $\eta$ and $\eta'$ meson production with
the LHCb detector. The $\eta$ and $\eta'$ cross sections are measured
differentially in $p_{\rm T}$ in $pp$ collisions at $\sqrt{s}=5.02$ and $13~{\rm
TeV}$, and in $p{\rm Pb}$ collisions at $8.16\,{\rm TeV}$. The $\eta$ meson
cross sections are combined with previous LHCb measurements of the $\pi^0$
differential cross sections to calculate the $\eta/\pi^0$ cross section ratio.
Finally, the nuclear modification factors are reported as a function of $p_{\rm
T}$. Since LHCb collected a relatively small sample of unbiased $pp$ collisions
at $\sqrt{s}=8\,{\rm TeV}$, the $pp$ reference cross section is constructed by
interpolating between the $\sqrt{s}=5.02$ and $13~{\rm TeV}$ cross section
measurements. The measurements are performed in the center-of-mass rapidity
($y_{\rm c.m.}$) ranges $2.5<y_{\rm c.m.}<3.5$ and $-4.0<y_{\rm c.m.}<-3.0$,
which correspond to the overlapping portions of the $pp$ and $p{\rm Pb}$
fiducial regions. The center-of-mass rapidity is related to the lab-frame
rapidity $y_{\rm lab}$ by $y_{\rm c.m.}=y_{\rm lab}-0.465$ in $p{\rm Pb}$
collisions and $|y_{\rm c.m.}|=y_{\rm lab}$ in $pp$ collisions. The $\eta$ meson
measurements are performed for $1.5<p_{\rm T}<10.0\gev{}$, while the $\eta'$
meson measurements are performed for $3<p_{\rm T}<10\gev{}$ (natural units are
used throughout this article).

\section{Detector and data set}

The \lhcb detector is a single-arm forward spectrometer covering the
\mbox{pseudorapidity} range $2<\eta <5$, described in detail in
Refs.~\cite{LHCb-DP-2008-001,LHCb-DP-2014-002}. The detector includes a
high-precision tracking system consisting of a silicon-strip vertex detector
(VELO) surrounding the $pp$ interaction region, a large-area silicon-strip
detector located upstream of a dipole magnet, and three stations of
silicon-strip detectors and straw drift tubes placed downstream of the magnet.
Different types of charged hadrons are distinguished using information from two
ring-imaging Cherenkov detectors. Photons are reconstructed and identified by a
calorimeter system consisting of scintillating-pad (SPD) and preshower (PRS)
detectors, an electromagnetic (ECAL), and a hadronic (HCAL) calorimeter. Muons
are identified by a system composed of alternating layers of iron and multiwire
proportional chambers. Particularly important for this analysis is the ECAL,
which consists of alternating layers of lead and scintillator and has an energy
$E$ resolution of
$13.5\%/\sqrt{E/\gev{}}\oplus5.2\%\oplus(0.32\gev{})/E$~\cite{LHCb-DP-2020-001}.
Simulated data samples are used to model the detector response to $\eta$ and
$\eta'$ reconstruction. In the simulation, $p{\rm Pb}$ collisions are generated
using EPOS-LHC~\cite{Pierog:2013ria}, while $pp$ collisions are generated using
\textsc{Pythia}~\cite{Sjostrand:2007gs}. Decays of unstable particles are
described by \evtgen~\cite{Lange:2001uf}, and final-state radiation is generated
using \photos~\cite{davidson2015photos}. The interaction of the generated
particles with the detector, and its response, are implemented using the \geant
toolkit~\cite{Allison:2006ve,Agostinelli:2002hh} as described in
Ref.~\cite{LHCb-PROC-2011-006}.

Proton-lead collisions are recorded in two configurations: the forward
configuration in which the proton beam travels from the interaction region
toward the spectrometer, and the backward configuration in which the lead ion
beam travels toward the spectrometer. The forward configuration data are used to
perform measurements for positive $y_{\rm c.m.}$, and the backward data are used
for negative $y_{\rm c.m.}$. The $p{\rm Pb}$ data used in this analysis were
collected in 2016 and correspond to an integrated luminosity of $350\pm9~{\rm
\mu b}^{-1}$ ($364\pm9~{\rm \mu b}^{-1}$) in the forward (backward)
direction~\cite{LHCb-PAPER-2018-048,LHCb-PAPER-2014-047}. All $p{\rm Pb}$ events
are required to have at least one reconstructed track in the VELO. The $pp$ data
were collected in 2015 and correspond to an integrated luminosity of
$3.49\pm0.07~{\rm nb}^{-1}$ ($6.86\pm0.14~{\rm nb}^{-1}$) for
$\sqrt{s}=5.02~{\rm TeV}$ ($\sqrt{s}=13~{\rm TeV}$).

Signal $\eta$ candidates are reconstructed from pairs of photons. Photons are
reconstructed from energy deposits (clusters) in the ECAL. The ECAL clusters are
required to have $p_{\rm T}>500\,{\rm MeV}$ and must be far from the
extrapolated trajectories of all reconstructed tracks to exclude clusters
produced by charged particles. In addition, clusters produced by hadrons are
removed using a neural network classifier that takes as input the significance
of the cluster's distance to the nearest track, variables describing the ECAL
and PRS cluster shapes, and energy deposited in the SPD, PRS, ECAL, and HCAL.
Selected $\eta$ candidates with a diphoton mass $M(\gamma\gamma)$ satisfying
$500<M(\gamma\gamma)<600~{\rm MeV}$ are combined with pairs of tracks to form
the $\eta'$ candidates. The tracks are required to be of good quality and be
consistent with originating from the same primary collision vertex (PV). The
tracks also must be identified as charged pions and have $p_{\rm T}>400\,{\rm
MeV}$.

\section{Cross section determination}

\begin{figure}
    \centering
    \includegraphics[width=0.48\textwidth]{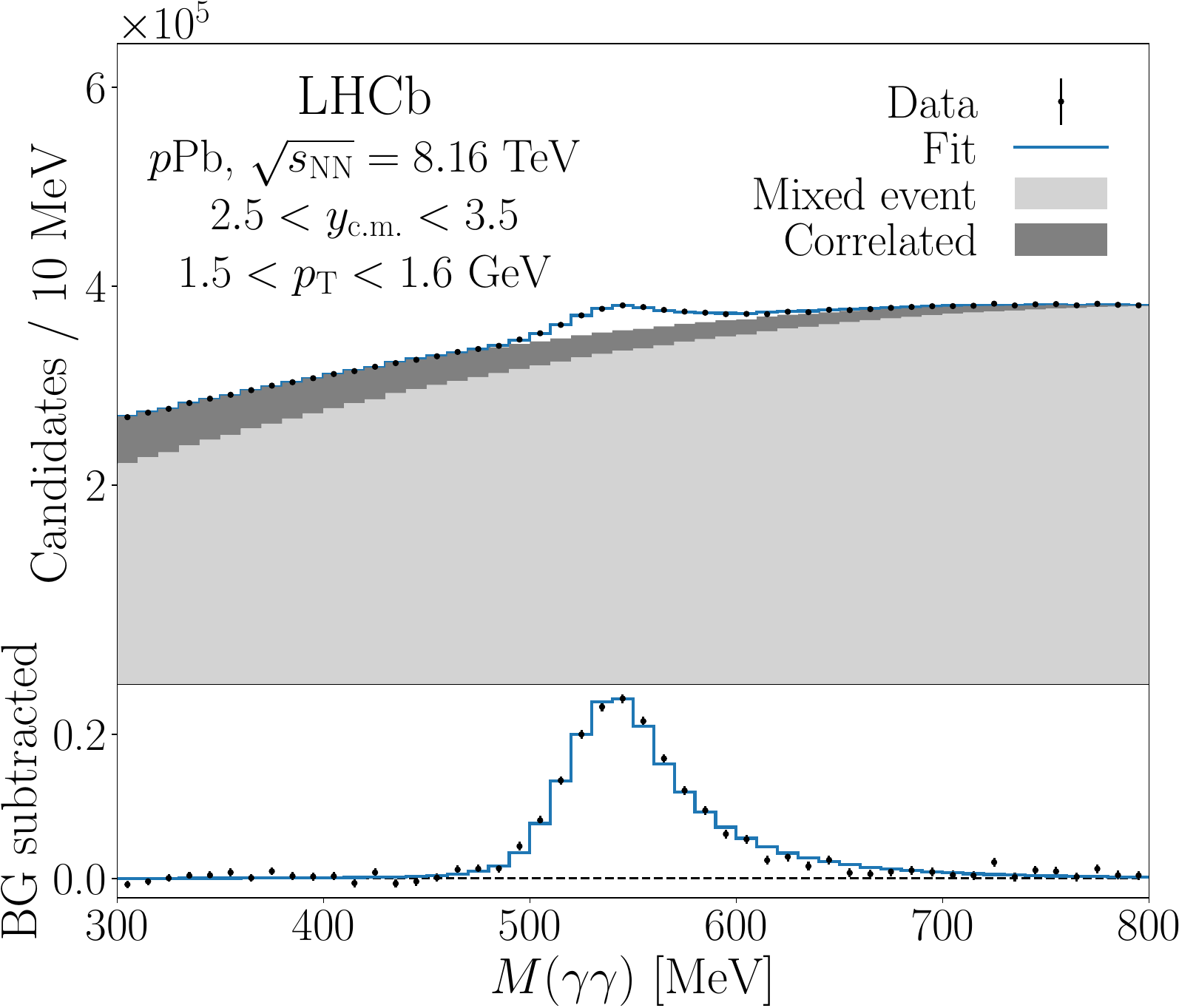}~~~
    \includegraphics[width=0.48\textwidth]{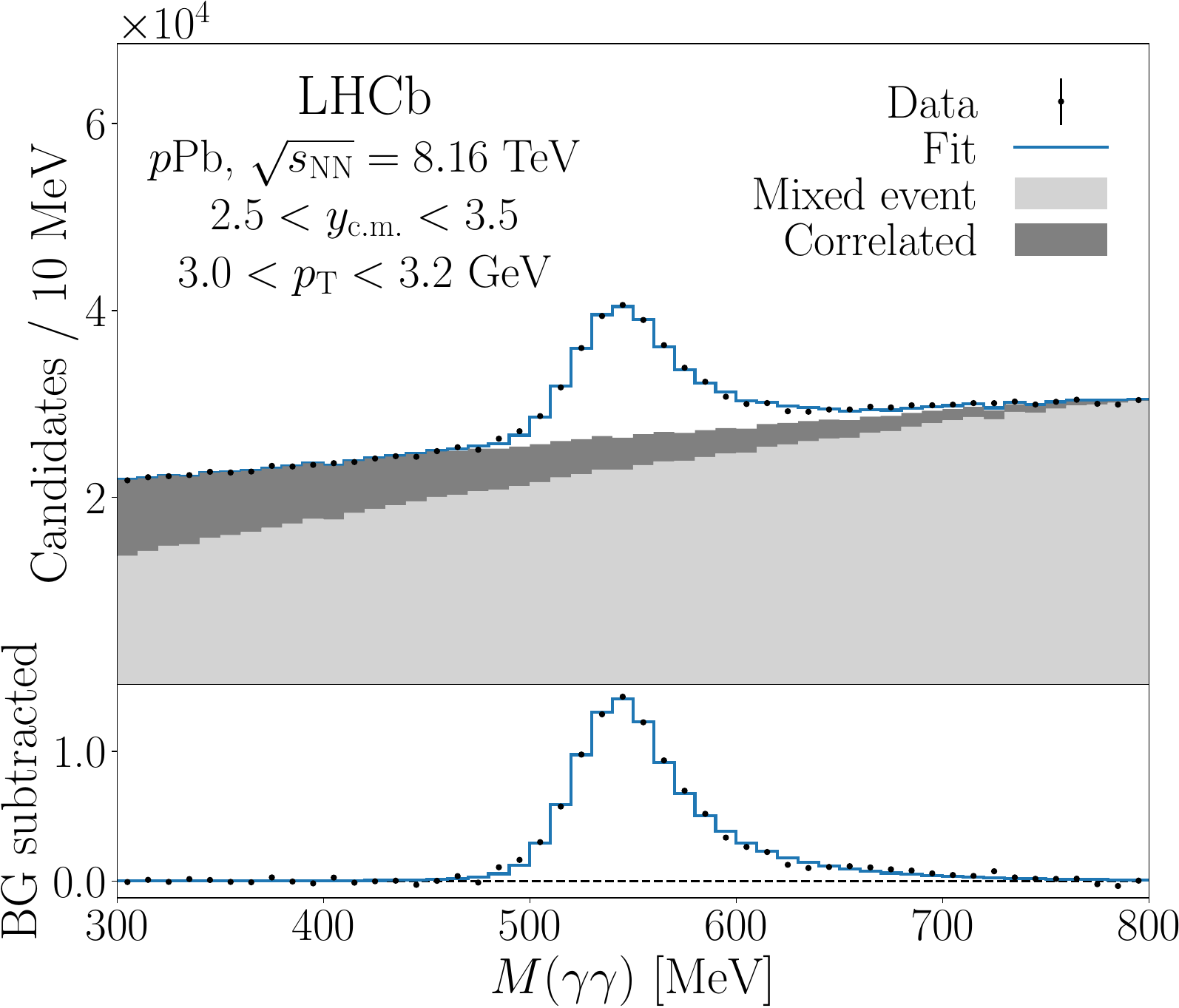}
    \caption{ Example reconstructed $M(\gamma\gamma)$ distributions in forward $p{\rm Pb}$
        collisions at \mbox{$\sqrt{s_{\rm NN}}=8.16~{\rm TeV}$} with
        \mbox{$2.5<y_{\rm c.m.}<3.5$}. Distributions are shown for (left)
        \mbox{$1.5<p_{\rm T}<1.6\gev{}$} and (right) \mbox{$3.0<p_{\rm
        T}<3.2\gev{}$}. Fit results are overlaid, including the mixed event and correlated combinatorial background templates.
        The lower panels show the
        background-subtracted mass distributions with the fit
        results overlaid.}
    \label{fig:etafits}
\end{figure}

The differential cross section in a particular $p_{\rm T}$ interval is given by
\begin{equation}
    \frac{d\sigma}{dp_{\rm T}}=\frac{1}{\mathcal{L}\times\mathcal{B}\times\epsilon(p_{\rm T})}\frac{N(p_{\rm T})}{\Delta p_{\rm T}},
    \label{eqn:xsec}
\end{equation}
where $\mathcal{L}$ is the integrated luminosity of the sample, $\mathcal{B}$ is
the branching ratio of the $\eta$ or $\eta'$ meson to the reconstructed final
state, and $\epsilon(p_{\rm T})$ is a correction factor that accounts for
detector inefficiencies and migration between $p_{\rm T}$ intervals due to
finite detector resolution. The signal yield in a given $p_{\rm T}$ interval is
denoted by $N(p_{\rm T})$, and $\Delta p_{\rm T}$ is the size of the $p_{\rm T}$
interval. The $\eta$ meson is studied using its decay to two photons and the
$\eta'$ meson is studied using its decay to $\pi^+\pi^-\eta$, which have
branching fractions of \mbox{$(39.36\pm0.18)\%$} and \mbox{$(42.5\pm0.5)\%$},
respectively~\cite{PDG2022}. The signal yields are determined using binned
maximum-likelihood fits to the mass spectra of reconstructed signal
candidates~\cite{iminuit,Dembinski:2022ios}. Example fit results from $p{\rm
Pb}$ collisions with \mbox{$2.5<y_{\rm c.m.}<3.5$} for \mbox{$1.5<p_{\rm
T}<1.6\gev{}$} and \mbox{$3.0<p_{\rm T}<3.2\gev{}$} are shown in
Fig.~\ref{fig:etafits}. The $\eta$ meson signal is modeled using a two-sided
Crystal Ball function~\cite{Skwarnicki:1986xj}. The parameters describing the
tails of the distribution are determined from fits to simulation, which are used
to impose Gaussian constraints in the fits to data. Most of the background
consists of random combinations of photons, which are modeled using an event
mixing technique in which reconstructed photons from different events are
combined. The mass distributions of the mixed-event photon pairs are used to
create background template histograms. In addition to these uncorrelated
combinations, an additional background source arises from the correlated
production of photons nearby in the detector, such as the combination of two
photons from the same jet. These photon pairs tend to be produced with small
opening angles. As a result, the background should be largest at the low edge of
the mass spectrum and monotonically decrease as a function of $M(\gamma\gamma)$.
This background is therefore modeled using a function of the form
\begin{equation}
    f(x)\propto 1-\left(\frac{x-m_0}{m_1-m_0}\right)^n,
    \label{eqn:corrbkg}
\end{equation}
where $x=M(\gamma\gamma)$, $m_0=300~{\rm MeV}$ is the low edge of the mass
spectra used in the fit, $m_1=800~{\rm MeV}$ is the high edge, and $n$ is a free
parameter. The combination of uncorrelated and correlated background models
provides a good description of the background in simulated samples of $\eta$
meson decays.

\begin{figure}
    \centering
    \includegraphics[width=0.48\textwidth]{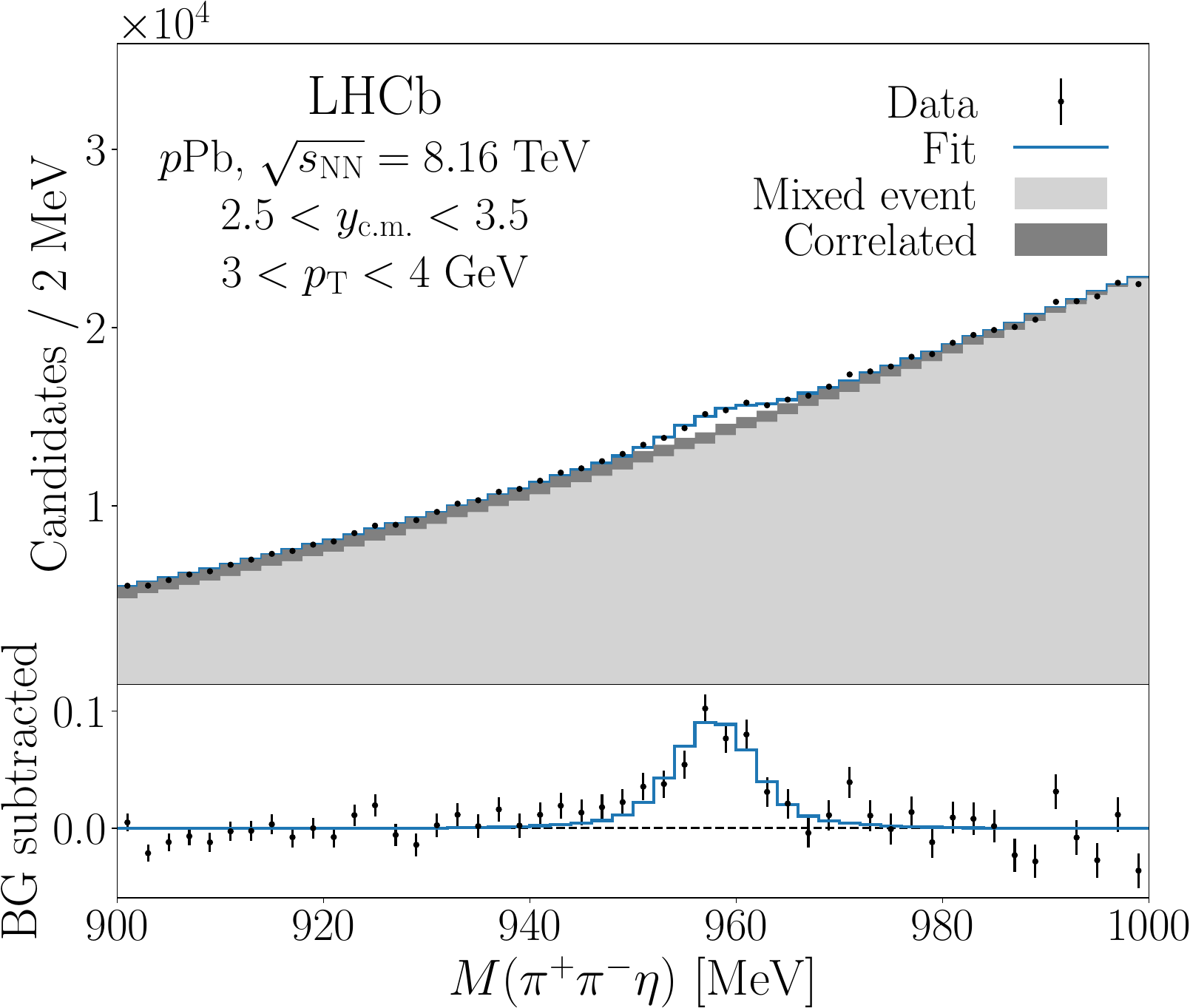}~~~
    \includegraphics[width=0.48\textwidth]{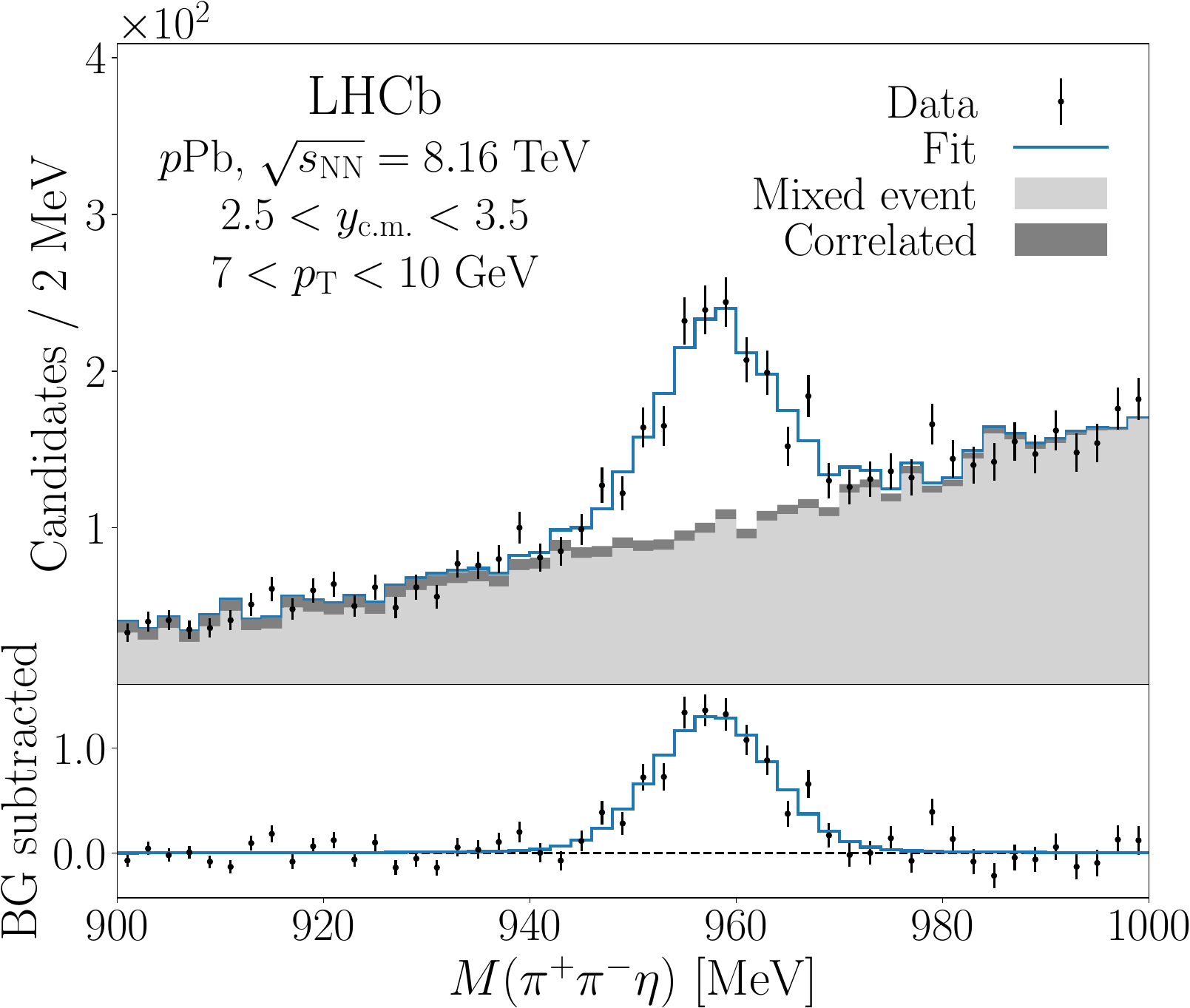}
    \caption{Example reconstructed $M(\pi^+\pi^-\eta)$ distributions in forward
        $p{\rm Pb}$ collisions at \mbox{$\sqrt{s_{\rm NN}}=8.16~{\rm TeV}$} with
        \mbox{$2.5<y_{\rm c.m.}<3.5$}. Distributions are shown for (left)
        \mbox{$3<p_{\rm T}<4\gev{}$} and (right) \mbox{$7<p_{\rm T}<10\gev{}$}.
        Fit results are overlaid, including the mixed event and correlated combinatorial background templates.
        The lower panels show the
        background-subtracted mass distributions with the fit results overlaid.}
    \label{fig:etapfits}
\end{figure}

The $\eta'$ candidate mass $M(\pi^+\pi^-\eta)$ is determined using a kinematic
fit with the decay vertex constrained to a PV and $M(\gamma\gamma)$ fixed to the
world-average $\eta$ meson mass~\cite{PDG2022}. Example fits from $p{\rm Pb}$
collisions with \mbox{$2.5<y_{\rm c.m.}<3.5$} for \mbox{$3<p_{\rm T}<4\gev{}$}
and \mbox{$7<p_{\rm T}<10\gev{}$} are shown in Fig.~\ref{fig:etapfits}. The
$\eta'$ meson signal is modeled using two Gaussian functions with a common mean.
The widths of the two Gaussian functions and their relative contributions to the
signal peak are determined from fits to simulation. The results of fits to
simulated events are used to impose Gaussian constraints in the fits to data.
The uncorrelated combinatorial background is modeled using mixed-event
combinations of dipion and $\eta$ candidates. Correlated combinatorial
background makes a much smaller relative contribution to the $\eta'$ mass
spectra than to that of the $\eta$ meson and is modeled using the functional
form of Eq.~\ref{eqn:corrbkg} with $x=M(\pi^+\pi^-\eta)$ with $n=1$,
$m_0=900~{\rm MeV}$, and $m_1=1000~{\rm MeV}$.

The signal yields are corrected for the effects of the detector response using
simulation. The correction factors $\epsilon$ are determined using an iterative
unfolding procedure. First, correction factors are calculated for each $p_{\rm
T}$ interval. Hagedorn functions are then fit to the corrected $p_{\rm T}$
spectra in data and the true $p_{\rm T}$ spectra in
simulation~\cite{PHENIX:2009gyd}. The ratio of these distributions is used to
weight the true signal $p_{\rm T}$ spectrum in simulation so that the true
$p_{\rm T}$ spectrum of the simulated data matches the $p_{\rm T}$ spectrum of
the corrected data. The procedure is then repeated using the weighted simulated
data sample. For the $\eta$ meson spectra, the procedure consistently converges
after three iterations. Since the momenta of the $\eta'$ candidates are carried
partially by charged particles, these candidates are reconstructed with better
momentum resolution than the $\eta$ candidates. Furthermore, the $\eta'$
measurement is performed in $p_{\rm T}$ intervals that are much larger than the
$p_{\rm T}$ resolution. Consequently, only one iteration is used to determine
the $\eta'$ correction factors.

The $\eta$ and $\eta'$ reconstruction efficiencies are calibrated using data.
The ECAL cluster reconstruction efficiency is measured using a tag-and-probe
method with photons that convert to an $e^+e^-$ pair in the detector material
upstream of the LHCb magnetic-field region~\cite{LHCb-DP-2019-003}. These
converted photons are reconstructed as pairs of tracks. The tag electron must be
matched to a cluster in the ECAL and identified as an electron, and the cluster
efficiency is the fraction of probe electrons matched to an ECAL cluster. The
photon identification efficiency is studied using $\pi^0$ decays to two photons,
where one photon is a converted photon. The $\pi^0$ yields are extracted using a
maximum-likelihood fit to the diphoton mass spectrum following
Ref.~\cite{LHCb-PAPER-2021-053}. The photon identification efficiency is then
the fraction of $\pi^0$ mesons for which the ECAL photon passes the photon
identification criteria. The charged pion reconstruction and identification
performance is studied using a tag-and-probe method with $\KS$ decays to
$\pi^+\pi^-$ pairs.

The $\eta$ and $\eta'$ differential cross sections in $pp$ collisions at
$8.16\,{\rm TeV}$ are estimated by interpolating between the measured $pp$ cross
sections at $\sqrt{s}=5.02$ and $13~{\rm TeV}$. The interpolation is performed
independently in each $p_{\rm T}$ interval. The cross section is interpolated
using the functional form $\sigma(\sqrt{s})=a\left(\sqrt{s}\right)^b$. This
procedure is found to give the correct yields to within $1\%$ in simulated $pp$
collisions at $\sqrt{s}=8.16~{\rm TeV}$ generated using \pythia{}. A linear
interpolation in $\sqrt{s}$ is also considered and provides less accurate
results in simulation.

\section{Systematic uncertainties}

\begin{table}[b]
    \centering

    \caption{Relative systematic and statistical uncertainties in
    $d\sigma^{\eta}/dp_{\rm T}$, $R^{\eta}_{p{\rm Pb}}$,
    $d\sigma^{\eta'}/dp_{\rm T}$, and $R^{\eta'}_{p{\rm Pb}}$, where the
    superscript denotes the hadron species. The uncertainties are reported in
    percent, and the ranges correspond to the minimum and maximum values of the
    associated uncertainties across all $p_{\rm T}$ intervals, $y_{\rm c.m.}$
    regions, and data sets. All sources of systematic uncertainty are
    approximately fully correlated across $p_{\rm T}$ intervals.}

    \begin{tabular}{l c c c c}
        \hline
        Source (\%) & $d\sigma^{\eta}/dp_{\rm T}$ & $R^{\eta}_{p{\rm Pb}}$ & $d\sigma^{\eta'}/dp_{\rm T}$ & $R^{\eta'}_{p{\rm Pb}}$ \\
        \hline
        Fit model           & $0.7 \mbox{--} 22.9$   & $0.5 \mbox{--} 12.3$  & $6.0 \mbox{--} 33.8$  & $2.5 \mbox{--} 26.4$ \\
        Unfolding           & $0.1 \mbox{--} 2.9$    & $0.1 \mbox{--} 2.5$   & $0.1 \mbox{--} 0.3$   & $0.1 \mbox{--} 0.4$ \\
        Interpolation       & \dots                 & $0.5 \mbox{--} 7.5$   & \dots                & $0.1 \mbox{--} 9.1$ \\
        Material budget            & $8.0$                 & \dots                & $10.8$               & \dots \\
        Photon efficiency   & $2.5 \mbox{--} 4.5$    & $1.9 \mbox{--} 4.7$   & $5.8 \mbox{--} 10.5$  & $10.1 \mbox{--} 11.5$ \\
        Tracking efficiency & \dots                 & \dots                & $0.2 \mbox{--} 0.5$   & $0.2 \mbox{--} 0.4$ \\
        Luminosity          & $2.0 \mbox{--} 2.6$    & $2.2 \mbox{--} 2.3$   & $2.0 \mbox{--} 2.6$   & $2.2 \mbox{--} 2.3$ \\
        \hline
        Total systematic    & $9.1 \mbox{--} 24.6$   & $3.4 \mbox{--} 14.8$  & $16.4 \mbox{--} 36.6$ & $10.8 \mbox{--} 29.1$ \\
        \hline
        Statistical         & $1.6 \mbox{--} 13.1$   & $2.7 \mbox{--} 10.6$  & $4.8 \mbox{--} 26.1$  & $6.8 \mbox{--} 16.4$ \\
        \hline
    \end{tabular}
    \label{tab:sys}
\end{table}

The systematic uncertainties are summarized in Table~\ref{tab:sys}. The largest
source of systematic uncertainty for the $\eta$ differential cross section in
most $p_{\rm T}$ regions is the detector material budget. The material budget is
proportional to the photon conversion probability, and uncertainties in the
material budget lead to uncertainties in the photon reconstruction efficiency.
The photon conversion probability has been measured with an uncertainty of
$4\%$. This uncertainty is fully correlated between photons, resulting in a
constant $8\%$ uncertainty on the differential cross section that fully cancels
in the $R_{p{\rm Pb}}$ measurement.

At low $p_{\rm T}$, the $\eta$ differential cross section uncertainty is
dominated by contributions from the mass fit model. A systematic uncertainty
associated with the choice of background model is estimated by replacing the
uncorrelated background model with an exponential function and taking the
difference with the default fit. An uncertainty associated with the choice of
signal model is estimated by extracting the yield by integrating the
background-subtracted mass spectra from the default fit in the range
$450<M(\gamma\gamma)<700~{\rm MeV}$. The difference in signal yields between the
default result and the background-subtracted result is taken as the signal model
uncertainty. The total fit-model uncertainty varies from greater than $20\%$ at
low $p_{\rm T}$ to less than around $5\%$ for $p_{\rm T}$ greater than about
$2\gev{}$. The fit-model uncertainty partially cancels in $R_{p{\rm Pb}}$.

An additional systematic uncertainty arises in the $R_{p{\rm Pb}}$ measurement
due to the $pp$ interpolation procedure. The interpolation uncertainty is
estimated by repeating the procedure using a linear interpolation model, as well
as a relative placement method~\cite{CMS:2015ved}. In the relative placement
procedure, placement factors are calculated using simulation assuming a linear
or a power-law dependence of the cross section on $\sqrt{s}$ and then applied to
data. The maximum variation of the interpolation result using these different
methods is taken as the interpolation uncertainty. The resulting uncertainty
varies between about $2\%$ and $5\%$.

Smaller sources of uncertainty come from unfolding, the luminosity estimates,
and the efficiency correction factors. The unfolding uncertainty arises from
differences in the $p_{\rm T}$ resolution of the $\eta$ candidates between data
and simulation, as well as differences between the underlying $p_{\rm T}$
distributions. The $p_{\rm T}$ resolution differences are estimated by comparing
the $\eta$ signal peak widths in data and simulation, and the resolution is
varied in the unfolding accordingly. The effect of differences in the underlying
$p_{\rm T}$ distribution is estimated by weighting the simulated data to vary
the initial $p_{\rm T}$ distribution. The resulting systematic uncertainty is
around $1\%$ or less in every $p_{\rm T}$ interval. The efficiency correction
uncertainty arises from the finite size of the simulated data samples and
results in global uncertainties of about $1-2\%$. The luminosity has been
measured in $pp$ collisions with a precision of $2\%$ and in $p{\rm Pb}$
collisions with a precision of $2.6\%$ in the forward configuration and $2.5\%$
in the backward configuration~\cite{LHCb-PAPER-2014-047}. The luminosity
uncertainty is $50\%$ correlated between datasets.

The fit model is the dominant source of uncertainty in the $\eta'$-related
measurements in most $p_{\rm T}$ intervals. The background model uncertainty is
estimated by repeating the fit using a polynomial function to model the
background and taking the difference with the default result. The signal model
uncertainty is estimated by integrating the background-subtracted mass spectrum
between $940$ and $980~{\rm MeV}$ and taking the difference with the default fit
result. The resulting total uncertainty varies between $6\%$ and about $34\%$.
This uncertainty partially cancels in $R_{p{\rm Pb}}$, resulting in
uncertainties of about $3\mbox{--}26\%$. The next largest source of uncertainty
is the material budget. The material budget uncertainty is the same as for the
$\eta$ measurement, with an additional $1.4\%$ uncertainty per charged pion in
the final state. The resulting total uncertainty is $10.8\%$ across the full
$p_{\rm T}$ range. This uncertainty fully cancels in the $R_{p{\rm Pb}}$
measurement. The photon reconstruction and identification correction uncertainty
is about $6\mbox{--}10\%$. Uncertainties associated with the track
reconstruction and particle identification efficiencies are determined by
varying the tracking efficiency calibration factors according to their
uncertainties. The resulting uncertainty is less than $1\%$ in every $p_{\rm T}$
interval. Additional uncertainties come from the luminosity determination,
unfolding, and interpolation. These sources of uncertainty are estimated using
the same methods used for the $\eta$ measurements and are subdominant.

\section{Results}

\begin{figure}[b]
    \centering
    \includegraphics[width=0.48\textwidth]{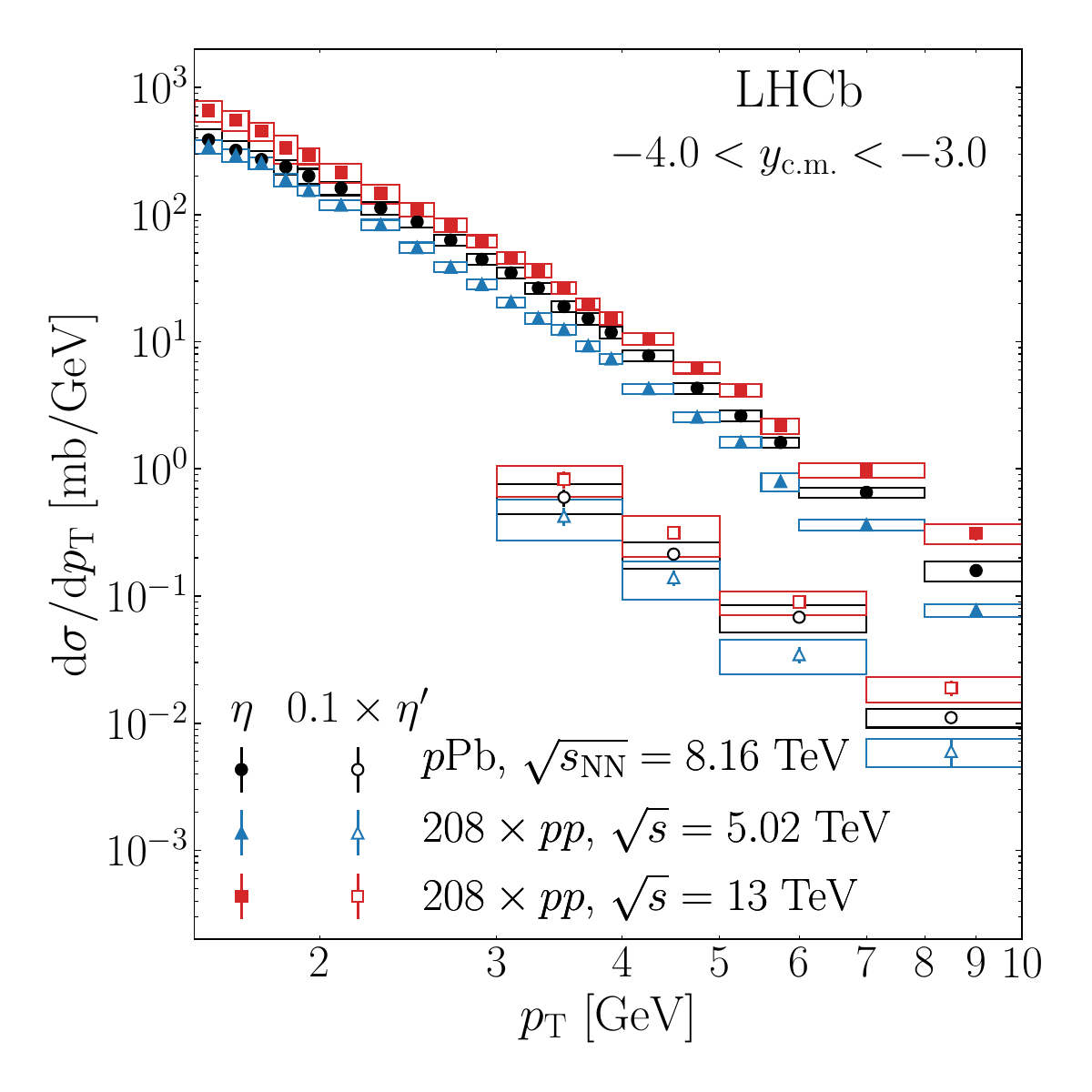}
    \includegraphics[width=0.48\textwidth]{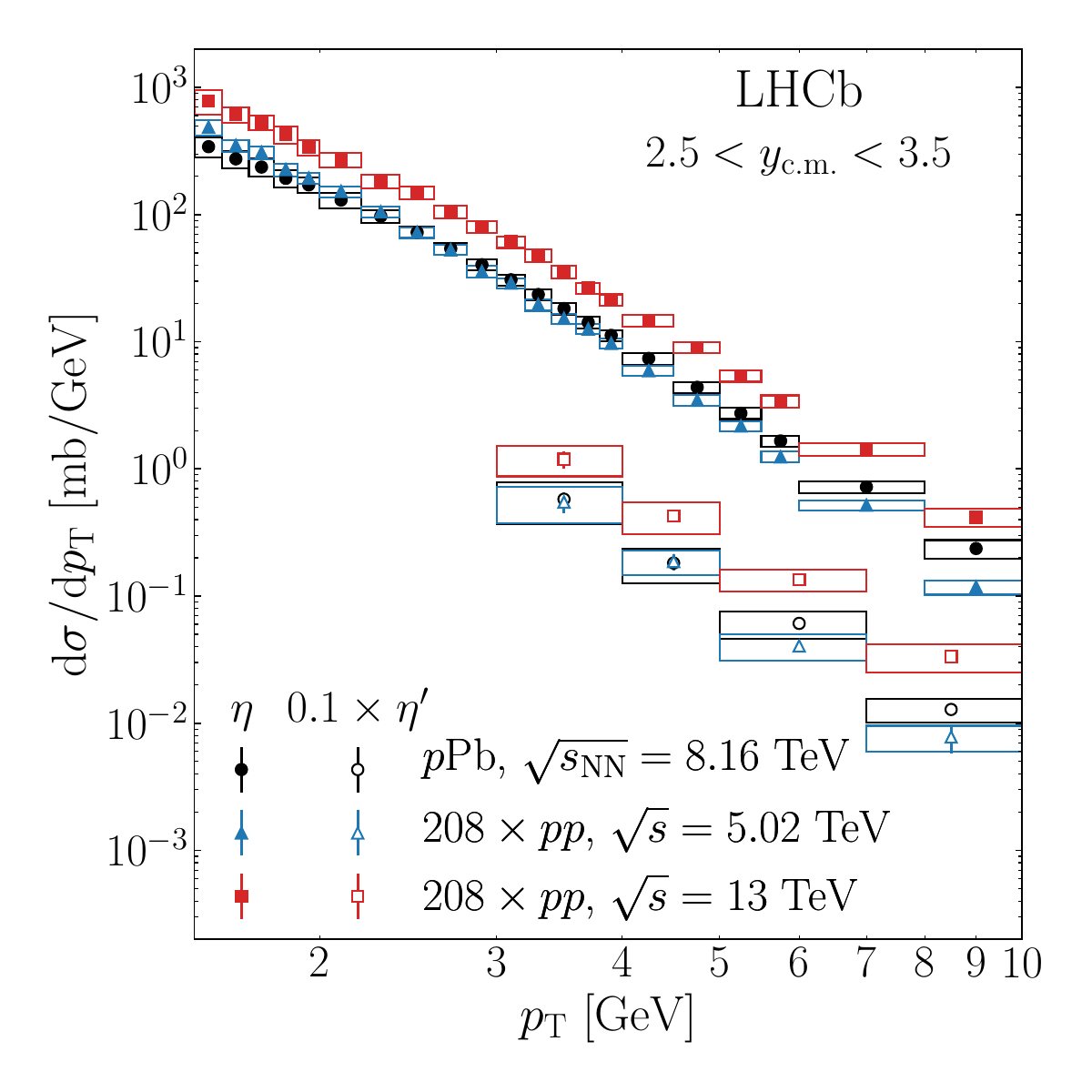}
    \caption{Measured $\eta$ and $\eta'$ differential cross sections in the
    (left) backward and (right) forward regions. The $pp$ cross sections are
    scaled by the atomic mass of the lead ion, $A=208$. The $\eta'$ cross
    sections are scaled down by a factor of $10$ for visual clarity. Statistical
    uncertainties are shown as error bars, while systematic uncertainties are
    shown as error boxes.}
    \label{fig:allxsec}
\end{figure}

\begin{figure}[h]
    \centering
    \includegraphics[width=0.75\textwidth]{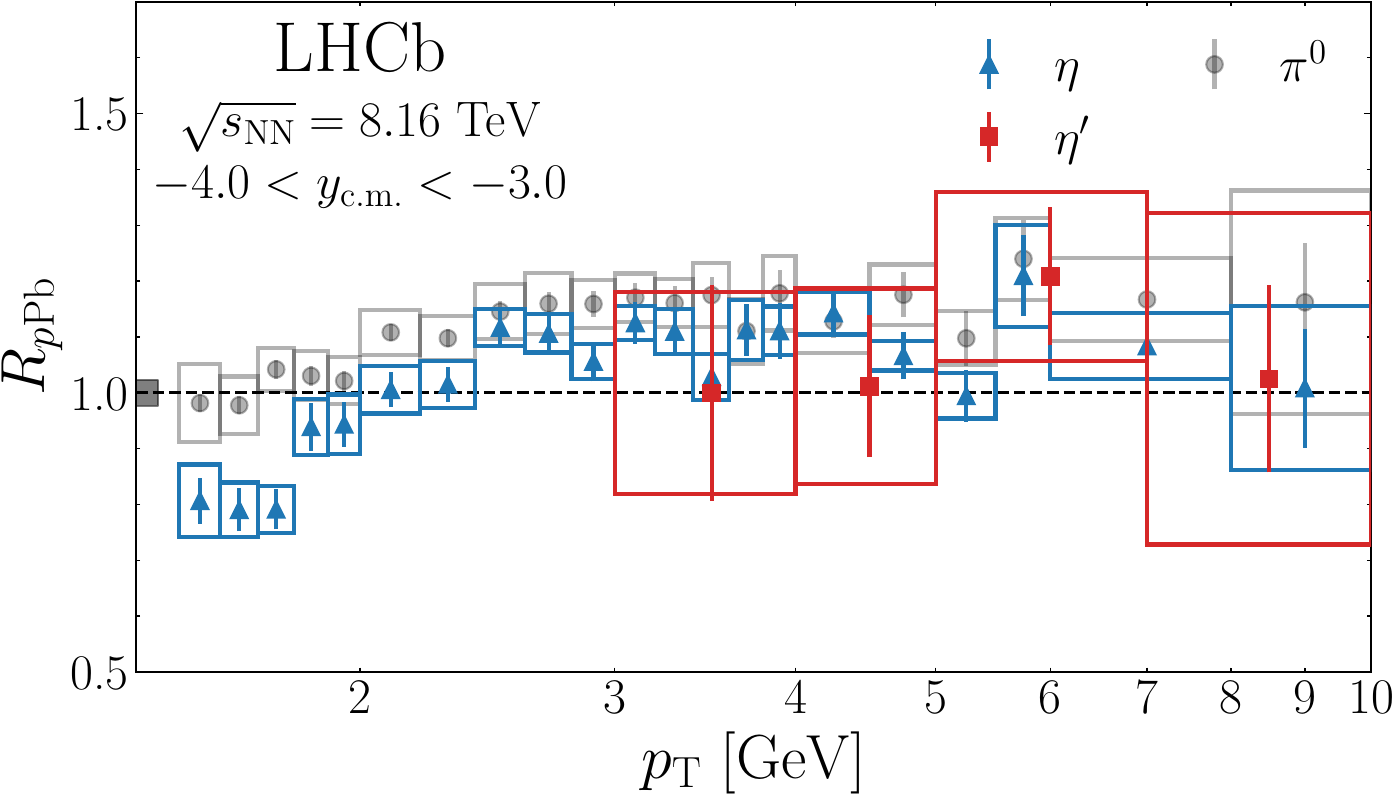}

    \smallskip
    \includegraphics[width=0.75\textwidth]{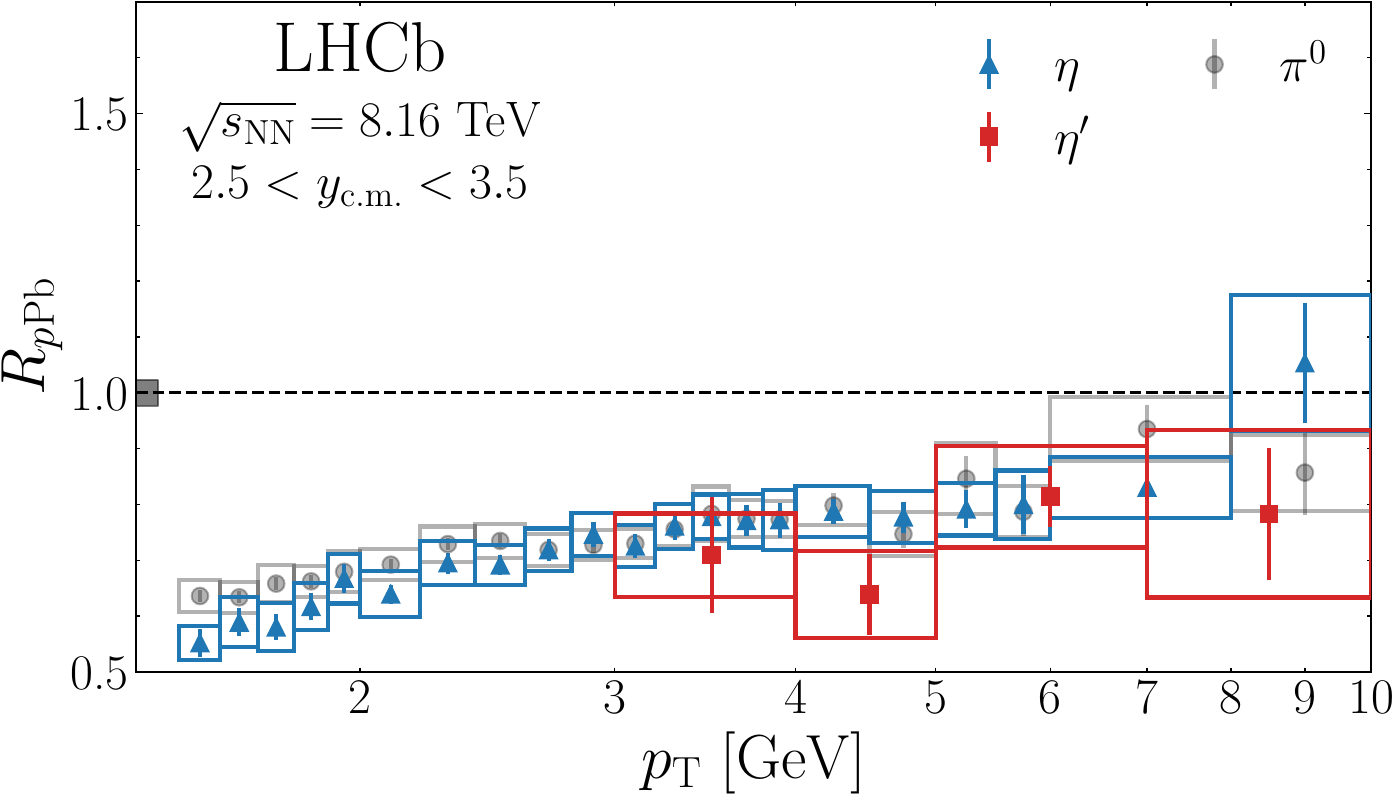}

    \caption{Measured $\eta$ and $\eta'$ nuclear modification factors in the
        (top) backward and (bottom) forward regions. Error bars show the
        statistical uncertainties, while the boxes show the systematic
        uncertainties except for the uncertainty associated with the luminosity,
        which is fully correlated between measurements. The luminosity
        uncertainty is shown as a dark gray shaded box. The $\eta$ and $\eta'$
        results are compared to the $\pi^0$ data from
        Ref.~\cite{LHCb-PAPER-2021-053}.}
    \label{fig:rpa}
\end{figure}

The measured $\eta$ and $\eta'$ differential cross sections are shown in
Fig.~\ref{fig:allxsec}. Results for the $\eta$ meson are tabulated in
Tables~\ref{tab:eta_xsec_pp2510}, \ref{tab:eta_xsec_pp6500}, and
\ref{tab:eta_xsec_pPb} (these and subsequent tables are in the appendix), and
the results for the $\eta'$ meson are tabulated in
Tables~\ref{tab:etap_xsec_pp2510}, \ref{tab:etap_xsec_pp6500}, and
\ref{tab:etap_xsec_pPb}. The differential cross sections are used to calculate
nuclear modification factors, which are shown in Fig.~\ref{fig:rpa} and
tabulated in Tables~\ref{tab:eta_rpa} and \ref{tab:etap_rpa}. In the forward
region, the $\eta$, $\pi^0$ and $\eta'$ results all agree where their fiducial
regions overlap. The observed suppression is consistent with the effects of
nuclear shadowing of the gluon density seen in global nPDF
analyses~\cite{Armesto:2006ph,Eskola:2021nhw,AbdulKhalek:2022fyi}. In the
backward region, the $\pi^0$ and $\eta$ measurements deviate at low $p_{\rm T}$
and converge for $p_{\rm T}>3\gev{}$. In this region, the $\pi^0$, $\eta$, and
$\eta'$ measurements all agree. The results show no significant evidence for
mass dependence of the nuclear modification factor of light neutral mesons.

\begin{figure}[h]
    \centering
    \includegraphics[width=0.48\textwidth]{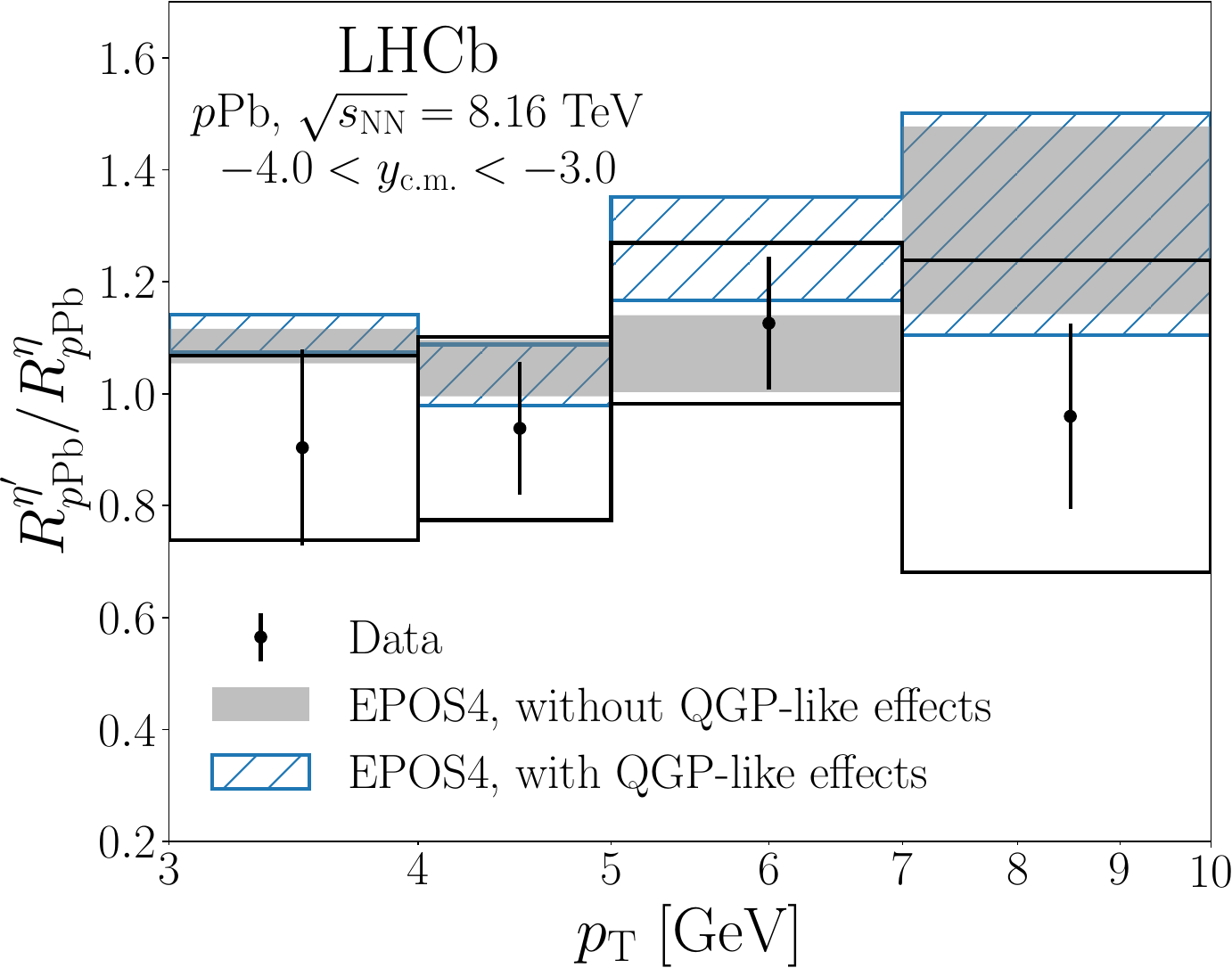}
    \includegraphics[width=0.48\textwidth]{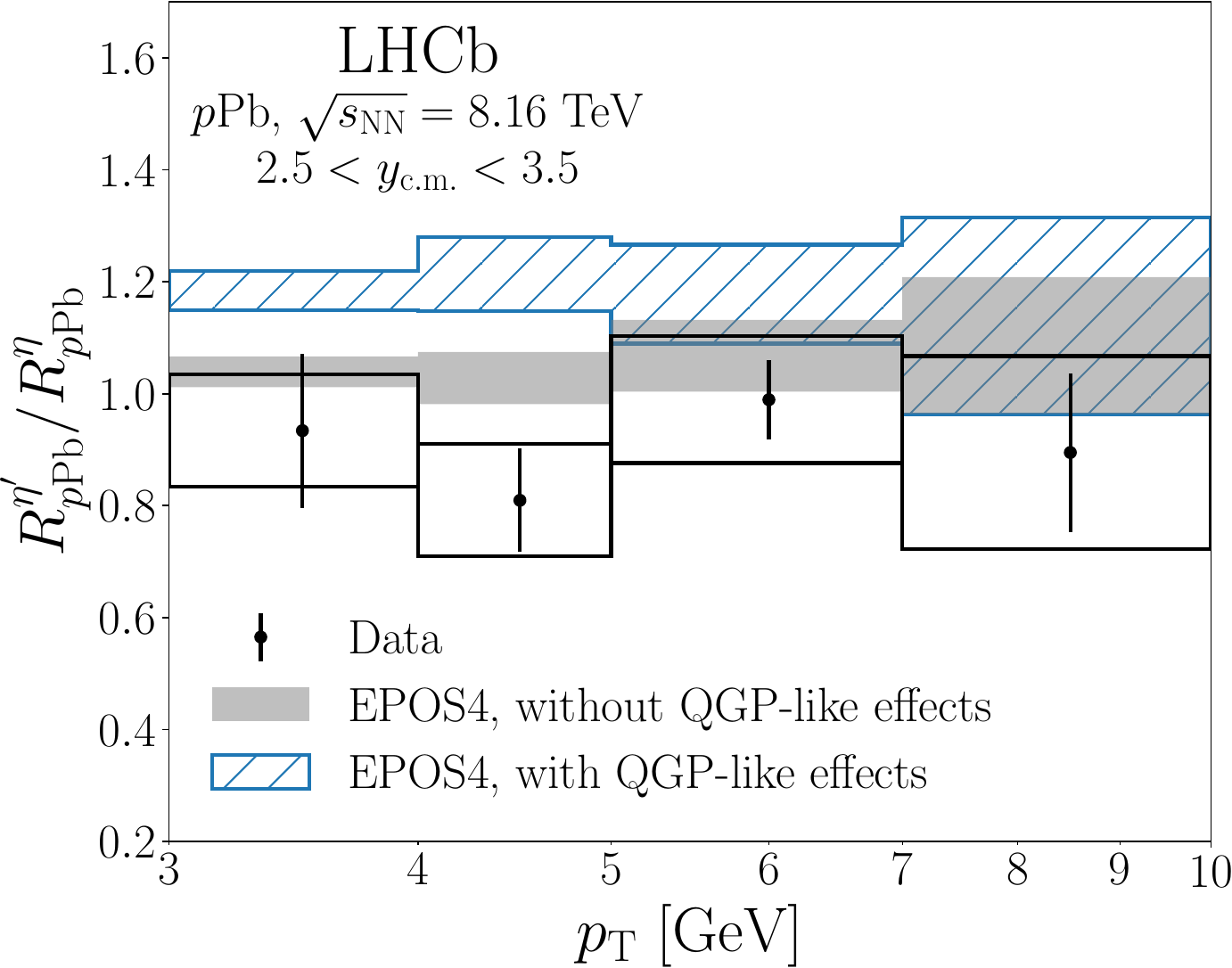}
    \caption{Measured $R^{\eta'}_{p{\rm Pb}}/R^{\eta}_{p{\rm Pb}}$ in the (left)
    backward and (right) forward regions. Error bars show the
    statistical uncertainties, while the boxes show the systematic
    uncertainties. The systematic uncertainties are approximately fully
    correlated in $p_{\rm T}$. The results are compared to predictions from
    EPOS4 with and without QGP-like effects. The shaded
    and hatched regions show the statistical 68\% confidence-level regions of
    the predictions.}
    \label{fig:etapetadata}
\end{figure}

The measured nuclear modification factor ratio $R^{\eta'}_{p{\rm
Pb}}/R^{\eta}_{p{\rm Pb}}$, where the superscript denotes the meson species of
the nuclear modification factor, is shown in Fig.~\ref{fig:etapetadata}. Since
the $R^{\eta'}_{p{\rm Pb}}$ and $R^{\eta}_{p{\rm Pb}}$ measurements are
performed in different $p_{\rm T}$ regions, the denominator of the ratio of
nuclear modification factors is constructed by fitting $R^{\eta}_{p{\rm Pb}}$
using a Gaussian process regression~\cite{GaussianProcess}. A systematic
uncertainty is obtained by repeating the fit while varying the $R^{\eta}_{p{\rm
Pb}}$ by its systematic uncertainties, assuming they are fully correlated in
$p_{\rm T}$. The resulting uncertainties are small relative to the
$R^{\eta'}_{p{\rm Pb}}$ uncertainties. The results are compared to predictions
from EPOS4 with and without QGP-like effects. The forward results show better
agreement with predictions without QGP-like effects, while the backward results
are in agreement with both sets of predictions.

\begin{figure}[h]
    \centering
    \includegraphics[width=0.48\textwidth]{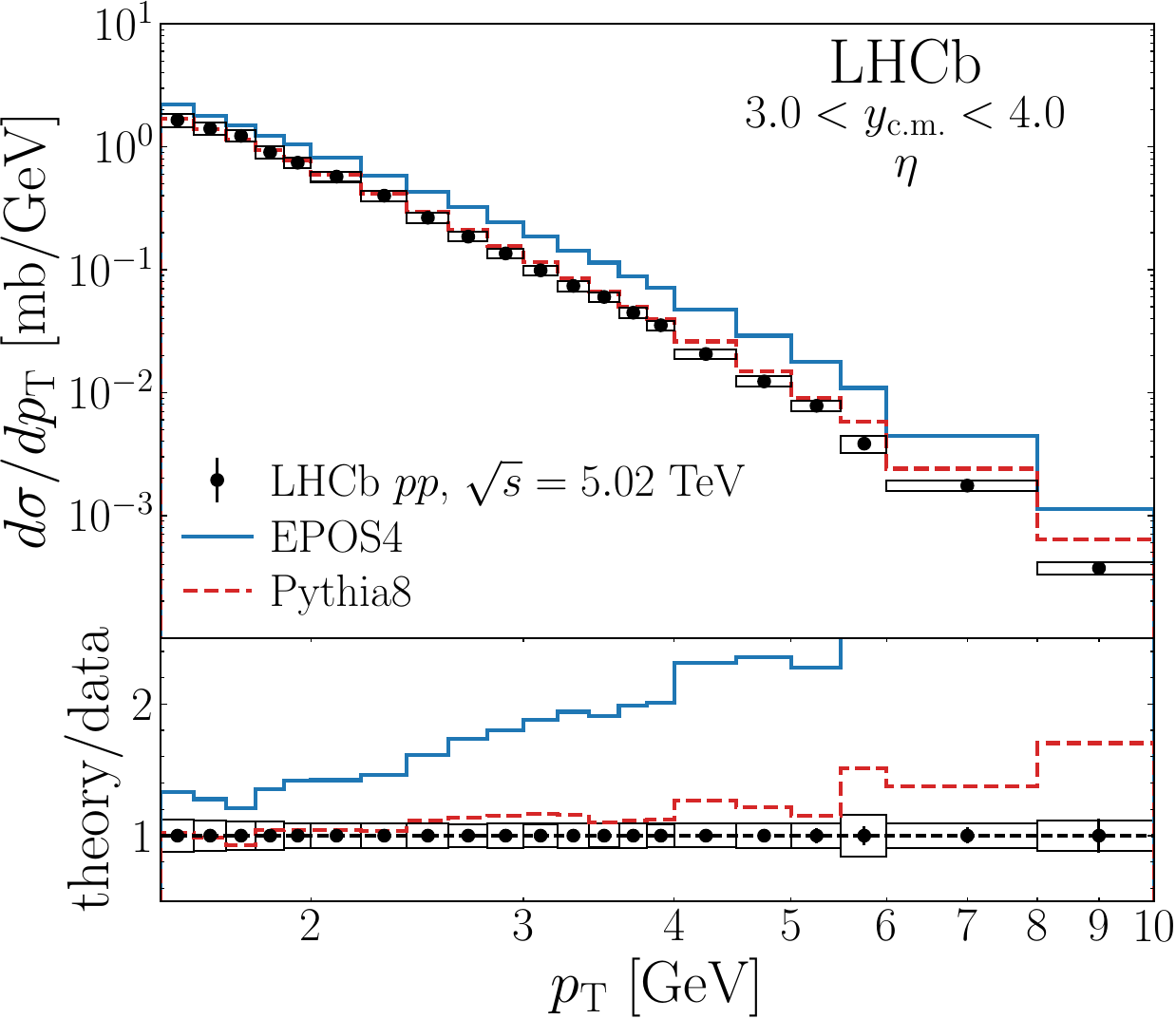}
    \includegraphics[width=0.48\textwidth]{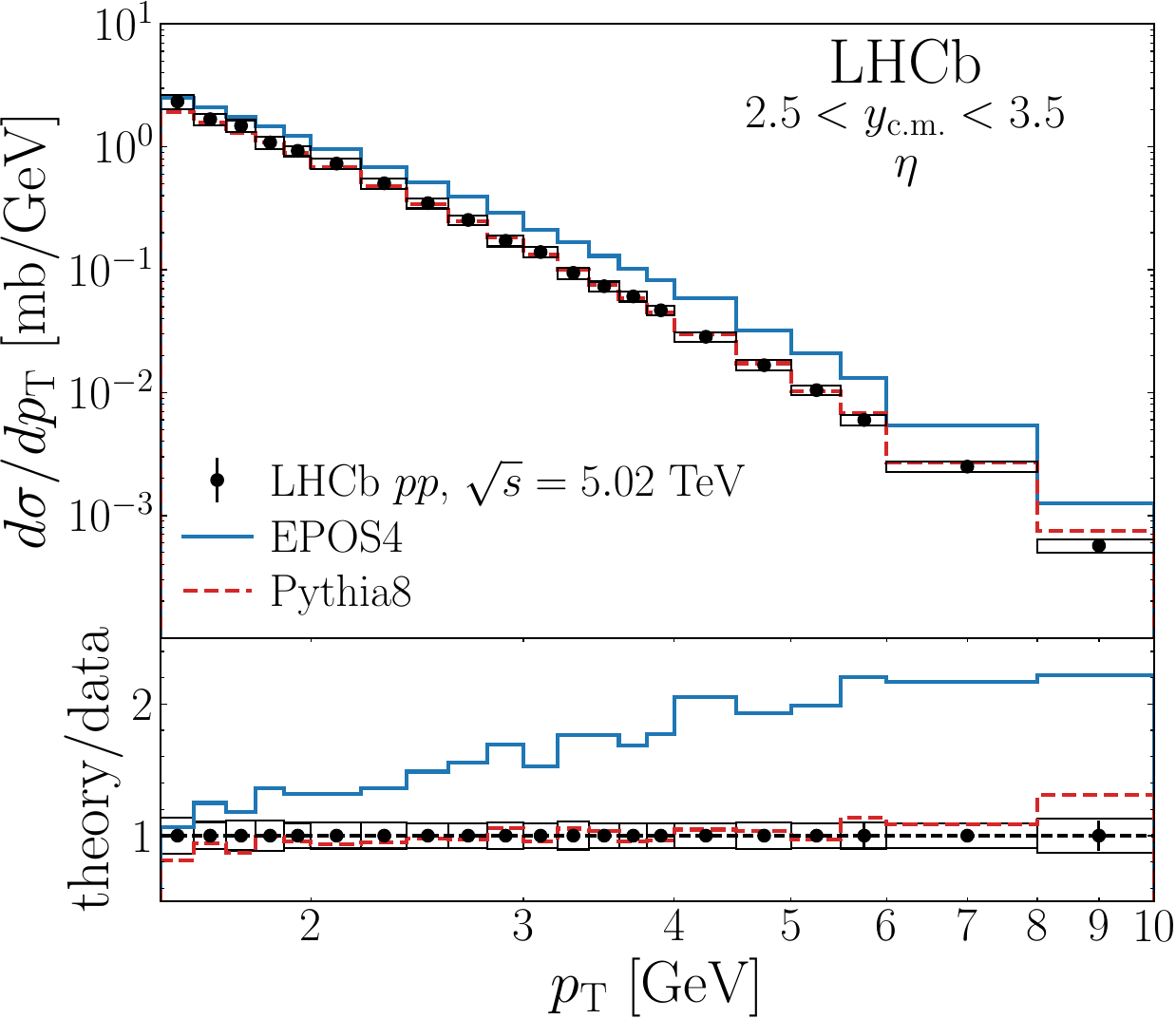}
    
    \includegraphics[width=0.48\textwidth]{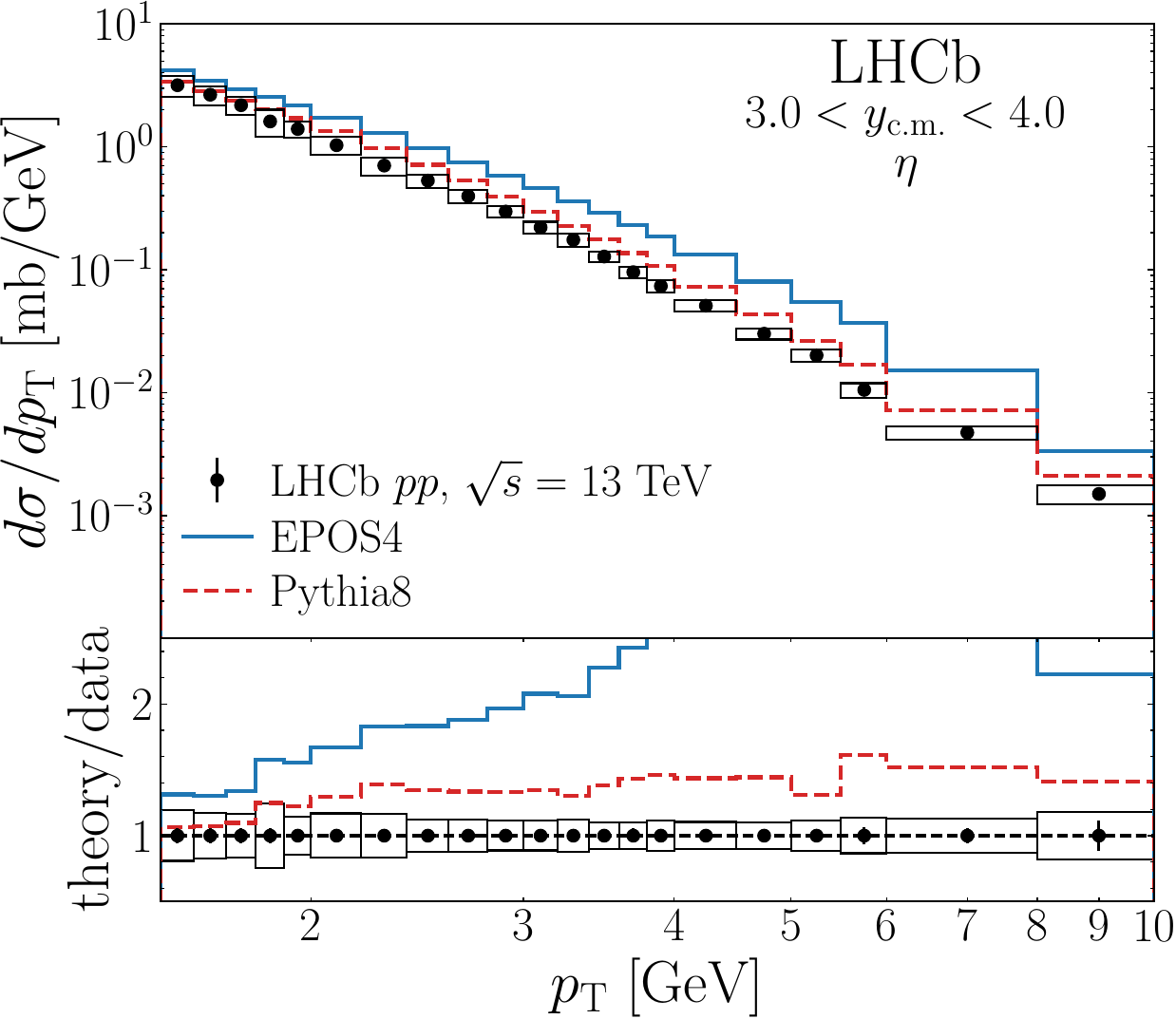}
    \includegraphics[width=0.48\textwidth]{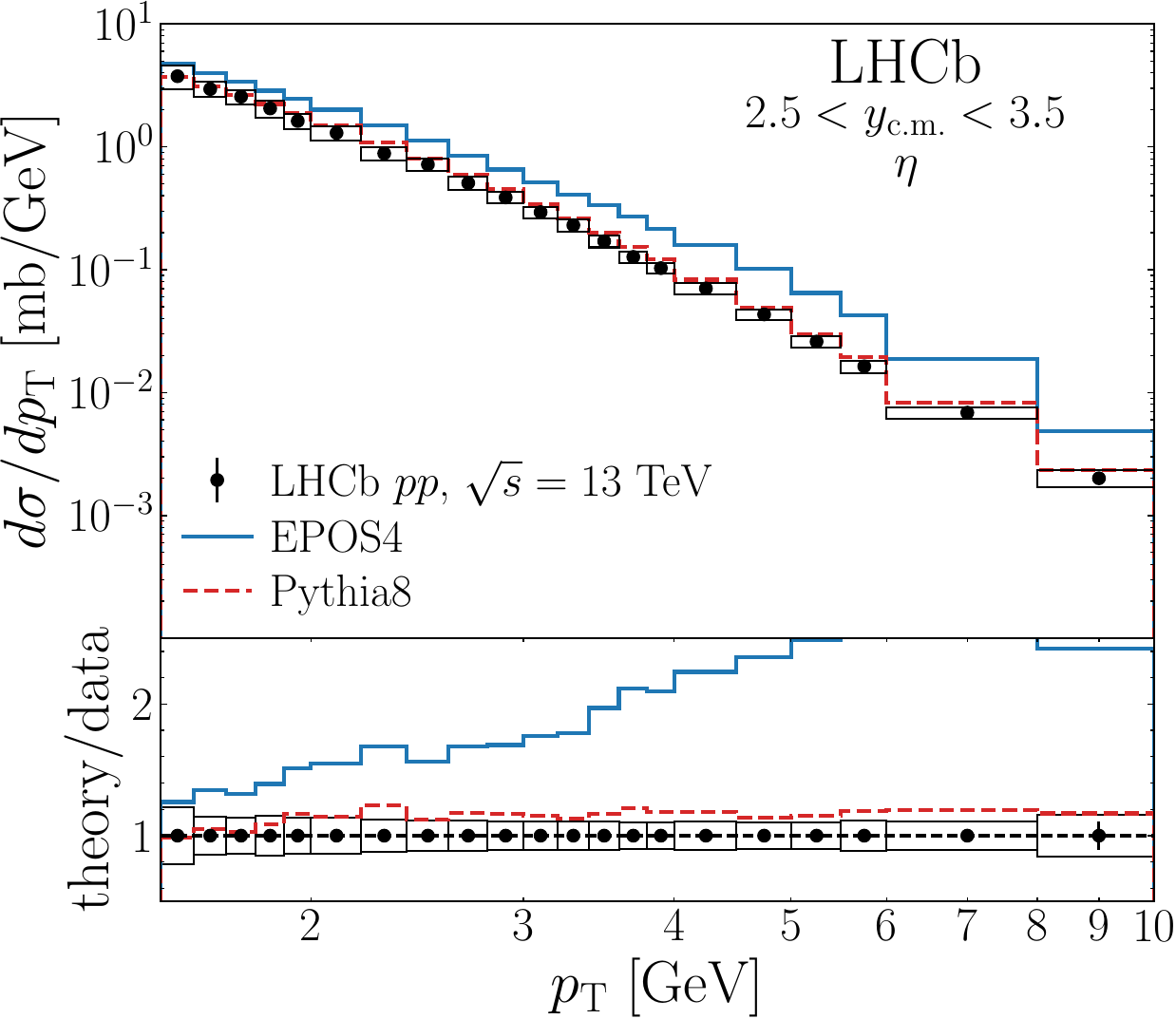}
    
    \includegraphics[width=0.48\textwidth]{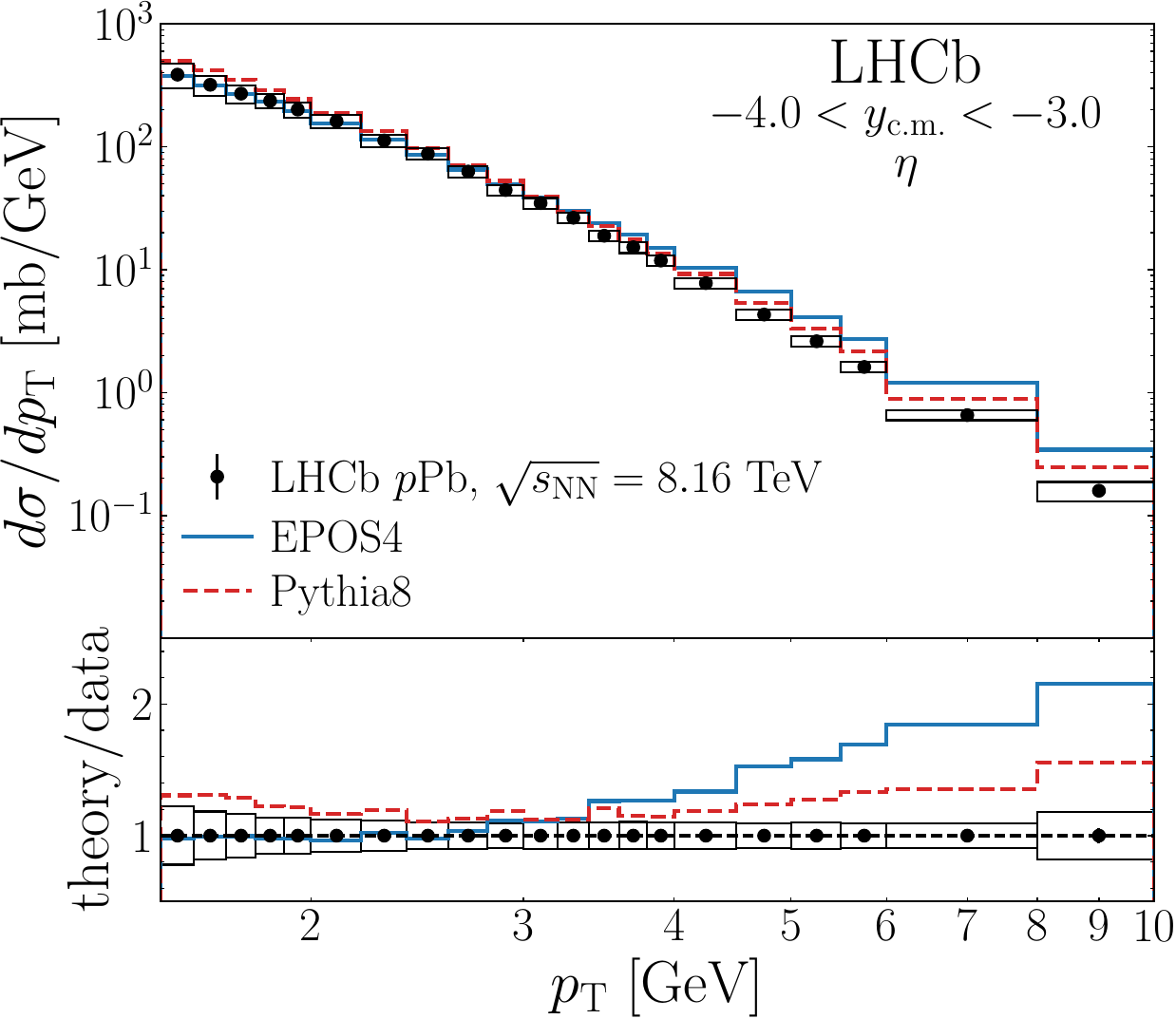}
    \includegraphics[width=0.48\textwidth]{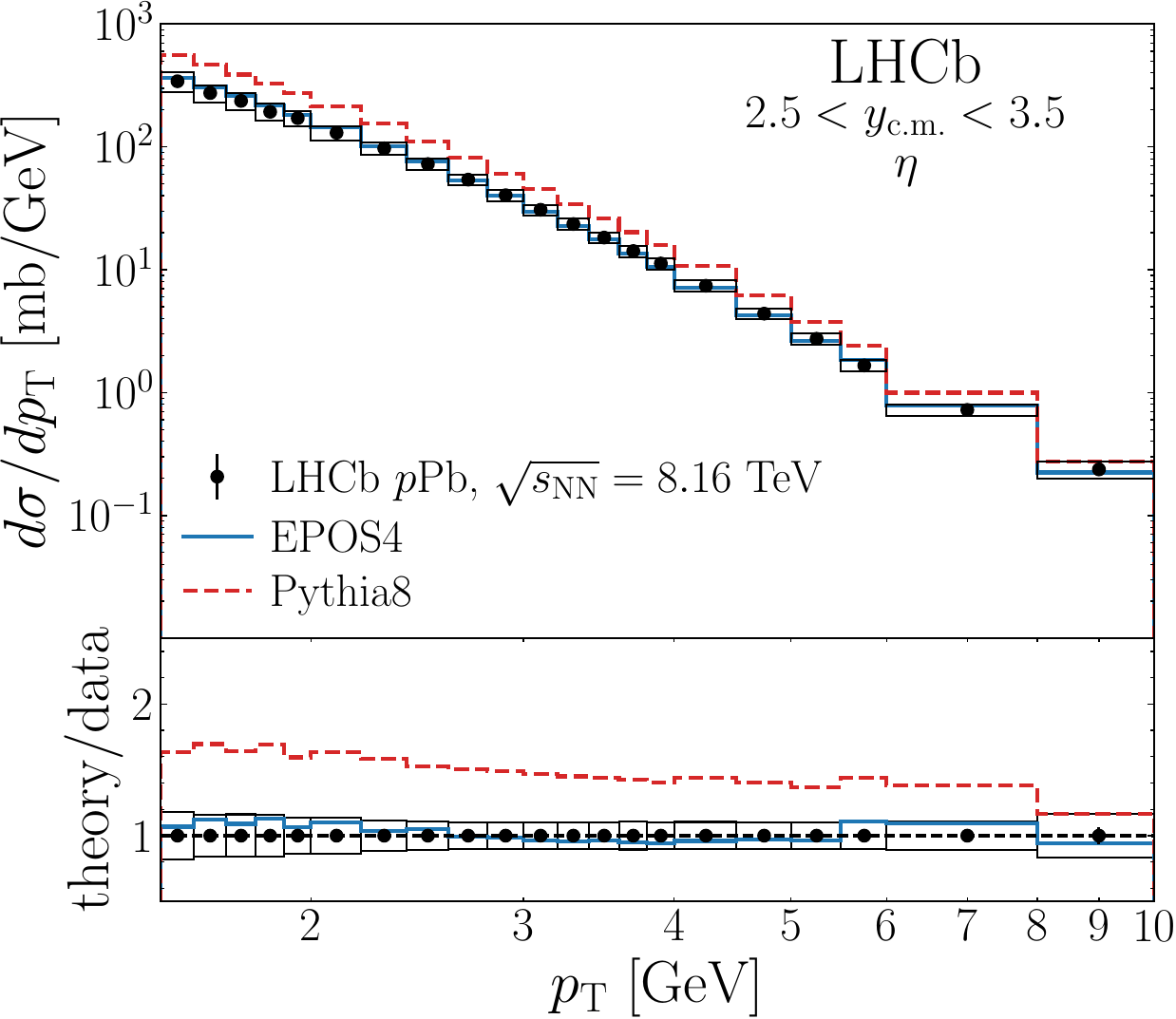}
    \caption{Measured $\eta$ differential cross sections in the (left) backward
    and (right) forward regions. Results are shown for $pp$ collisions at (top)
    $\sqrt{s}=5.02$ and (middle) $13~{\rm TeV}$ and (bottom) for $p{\rm Pb}$
    collisions at $\sqrt{s_{\rm NN}}=8.16~{\rm TeV}$. Results are compared to
    predictions from EPOS4 and \pythia{}8. The lower panels show the ratios of
    the predictions to the measured results. The statistical uncertainties are
    shown by error bars, while systematic uncertainties are shown by boxes.}
    \label{fig:etaxsec}
\end{figure}

\begin{figure}[h]
    \centering
    \includegraphics[width=0.48\textwidth]{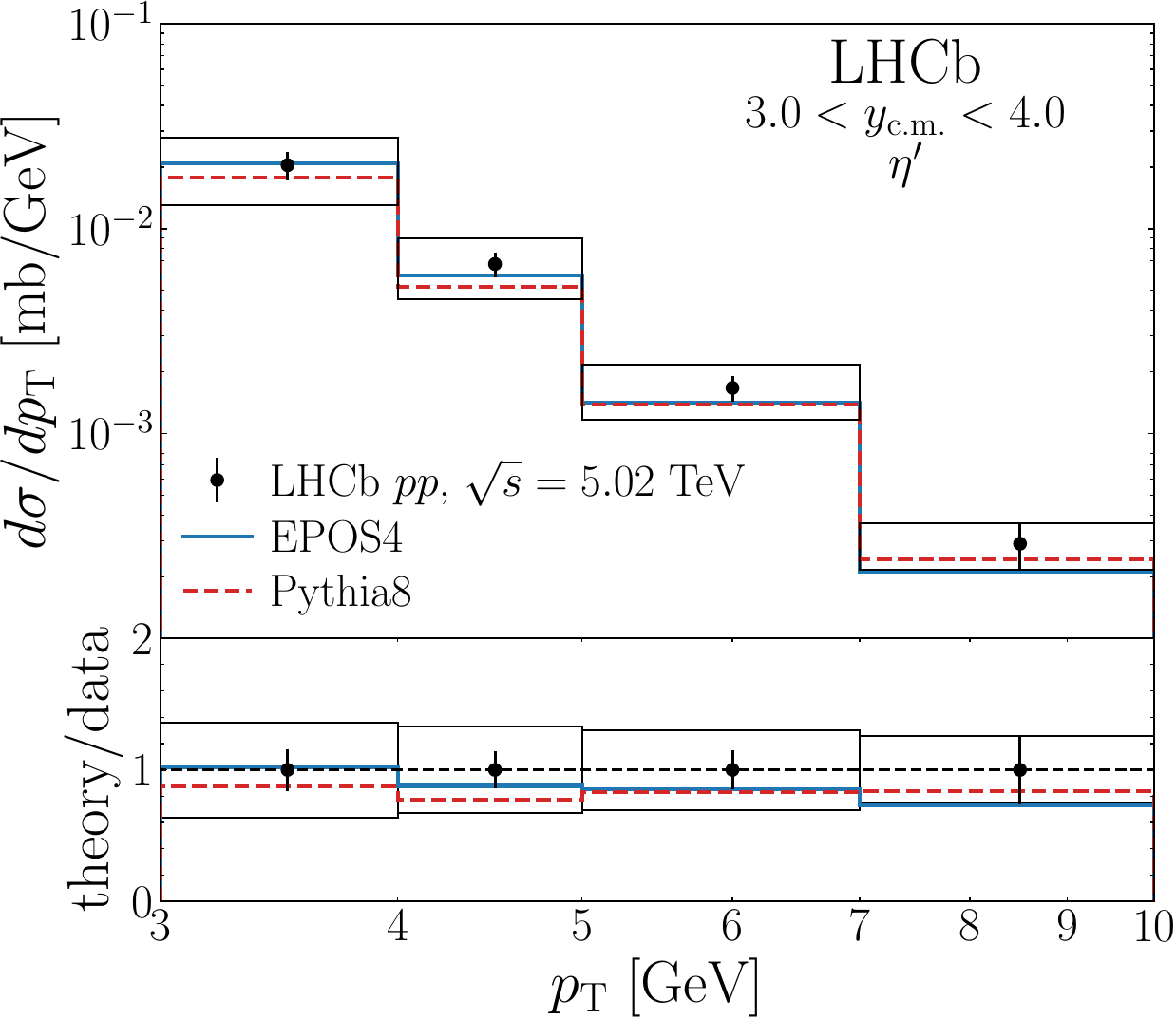}
    \includegraphics[width=0.48\textwidth]{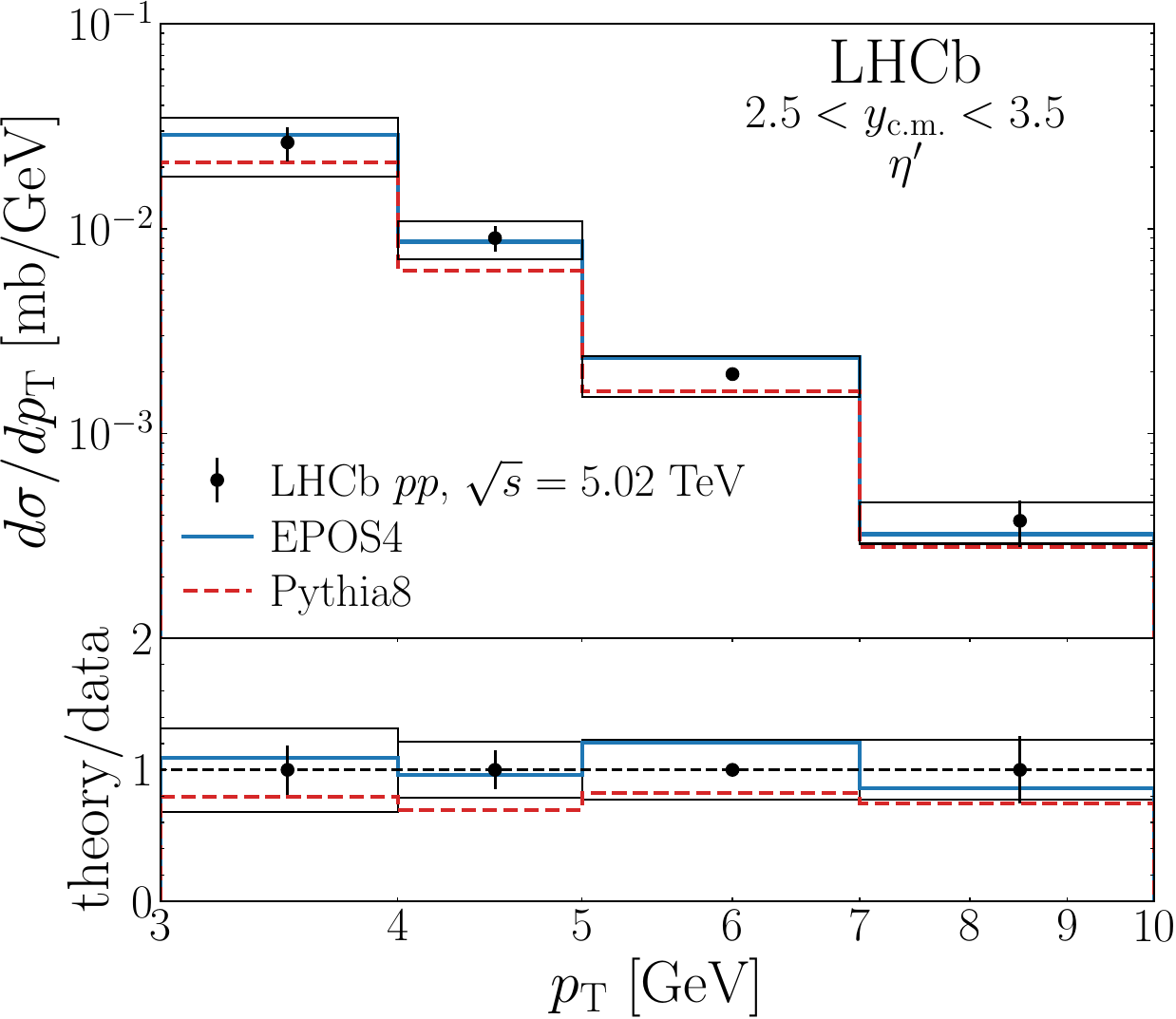}
    
    \includegraphics[width=0.48\textwidth]{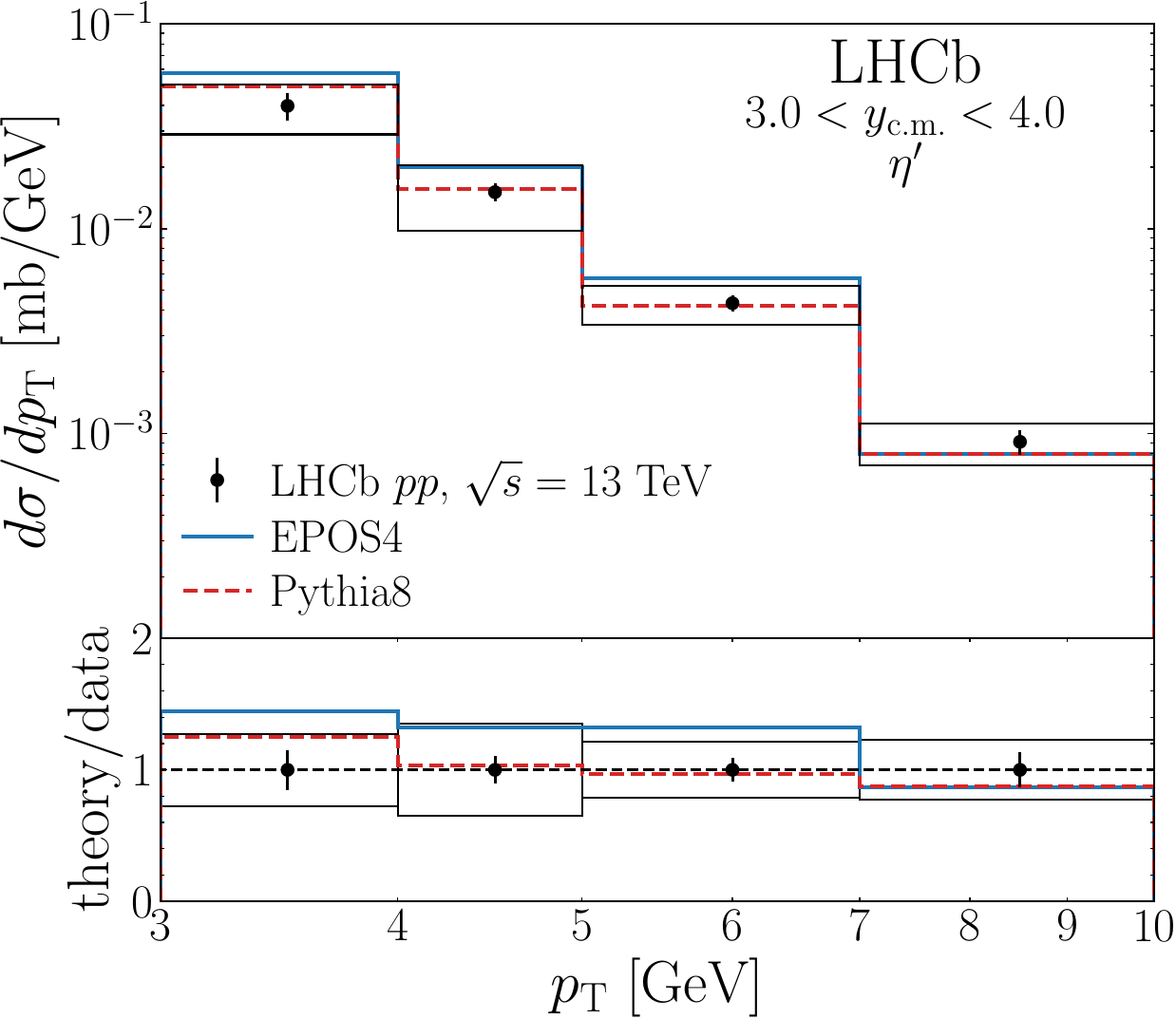}
    \includegraphics[width=0.48\textwidth]{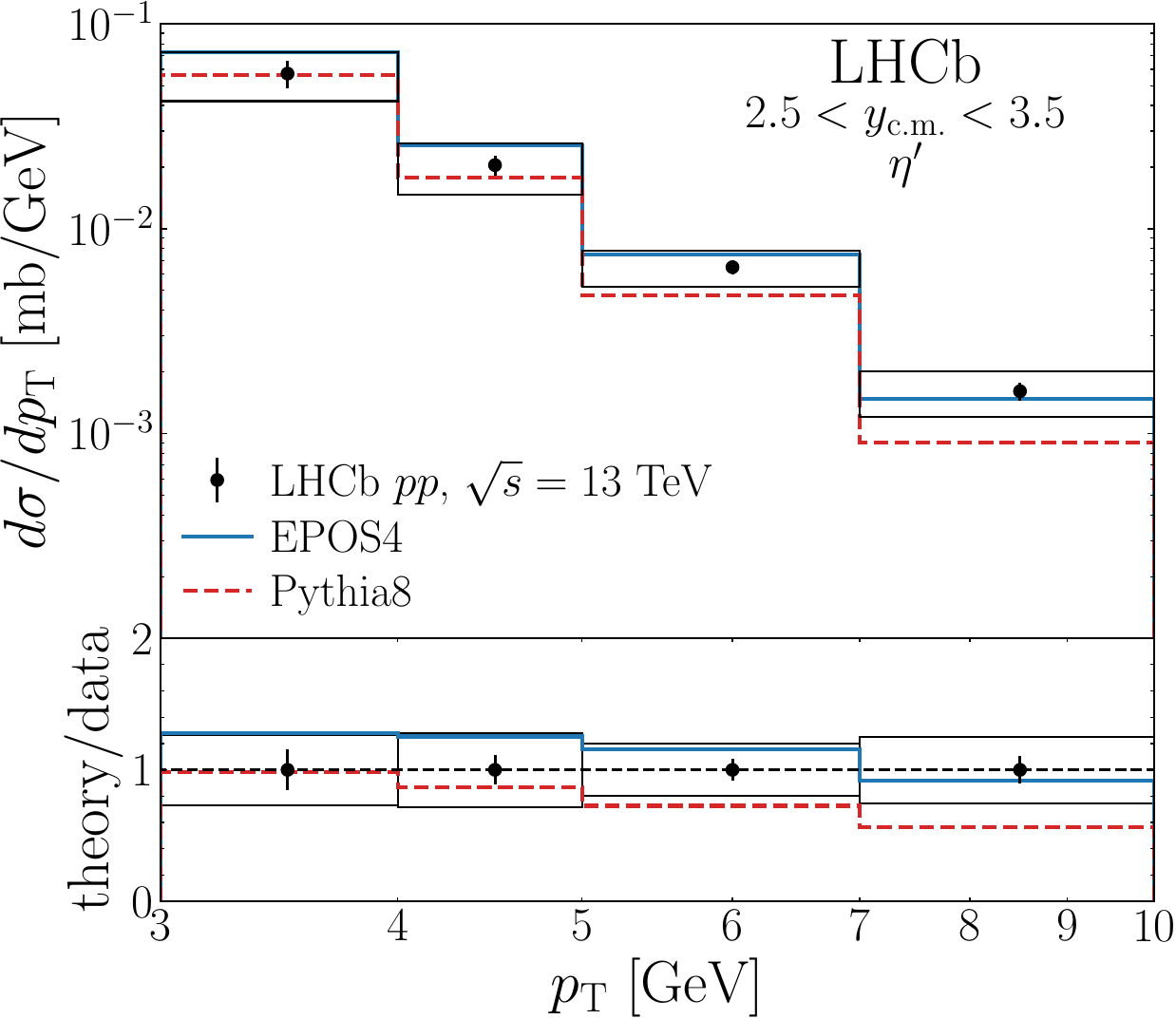}
    
    \includegraphics[width=0.48\textwidth]{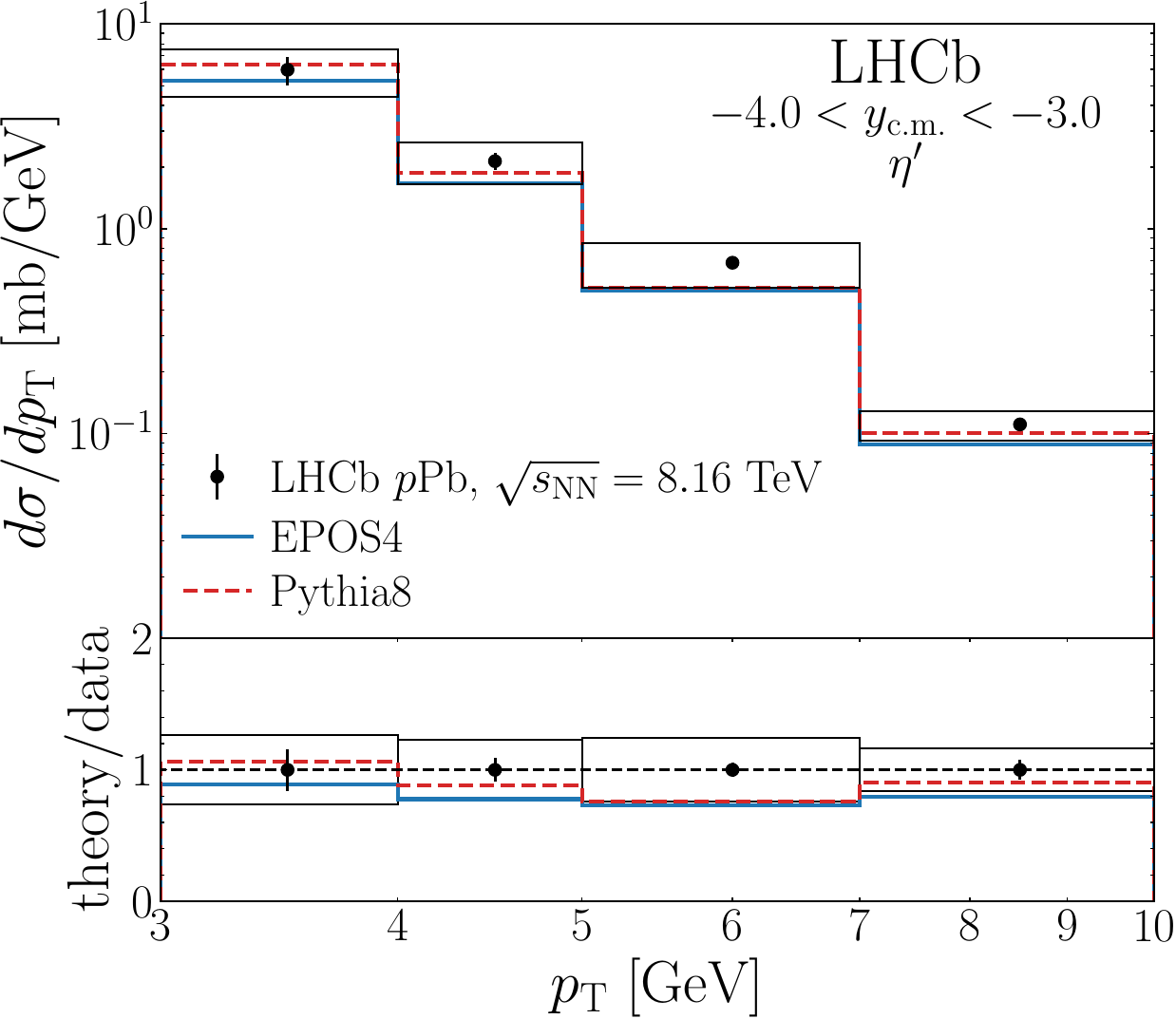}
    \includegraphics[width=0.48\textwidth]{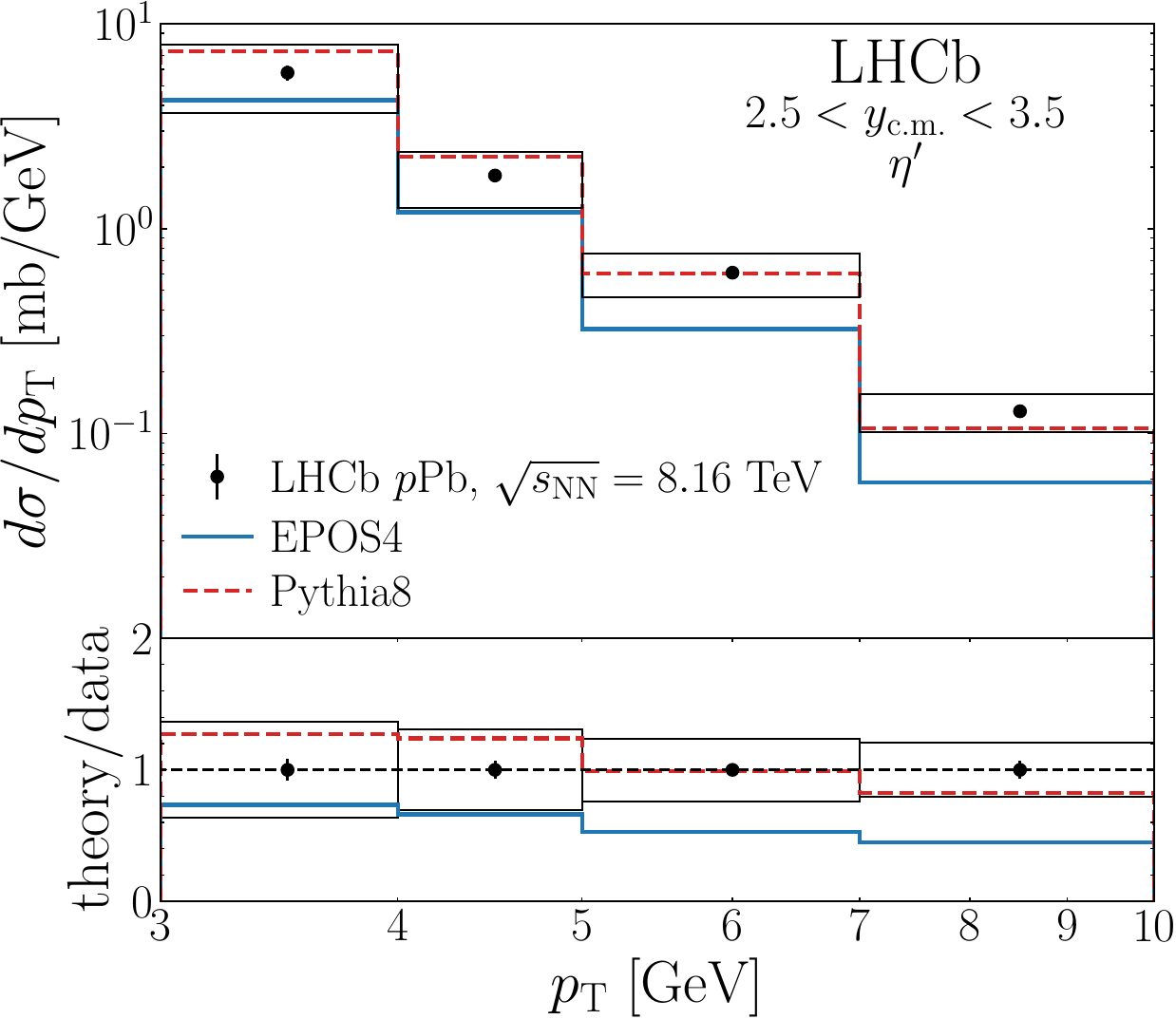}
    \caption{Measured $\eta'$ differential cross sections in the (left) backward
    and (right) forward regions. Results are shown for $pp$ collisions at (top)
    $\sqrt{s}=5.02$ and (middle) $13~{\rm TeV}$ and (bottom) for $p{\rm Pb}$
    collisions at $\sqrt{s_{\rm NN}}=8.16~{\rm TeV}$. Results are compared to
    predictions from EPOS4 and \pythia{}8. The lower panels show the ratios of
    the predictions to the measured results. The statistical uncertainties are
    shown by error bars, while systematic uncertainties are shown by boxes.}
    \label{fig:etapxsec}
\end{figure}

The differential cross sections are compared to calculations from \pythia{}8 and
EPOS4 in Figs.~\ref{fig:etaxsec} and \ref{fig:etapxsec}. The EPOS4 predictions
include the effects of hydrodynamic evolution and statistical hadronization for
both $pp$ and $p{\rm Pb}$ collisions and use a preliminary tune for $p{\rm Pb}$
collisions. The $p{\rm Pb}$ \pythia{}8 predictions are calculated by scaling
$pp$ predictions at $\sqrt{s}=8.16~{\rm TeV}$ by $A=208$. The \pythia{}8
predictions provide a good description of the $\eta$ results at
$\sqrt{s}=5.02~{\rm TeV}$ and at $\sqrt{s}=13~{\rm TeV}$ in the forward region,
but overestimate the $\eta$ meson yields at $\sqrt{s}=13~{\rm TeV}$ in the
backward region. The \pythia{}8 calculations do not use nPDFs or include other
nuclear effects, resulting in disagreement with the $p{\rm Pb}$ data. The EPOS4
predictions consistently overestimate the $\eta$ cross section in $pp$
collisions, but offer a good description of the backward $p{\rm Pb}$ results at
low $p_{\rm T}$ and the forward $p{\rm Pb}$ results over the entire $p_{\rm T}$
range. The $\eta'$ results generally agree with both the EPOS4 and \pythia{}8
predictions, although the measurement uncertainties are large.

\begin{figure}[h]
    \centering
    \includegraphics[width=0.48\textwidth]{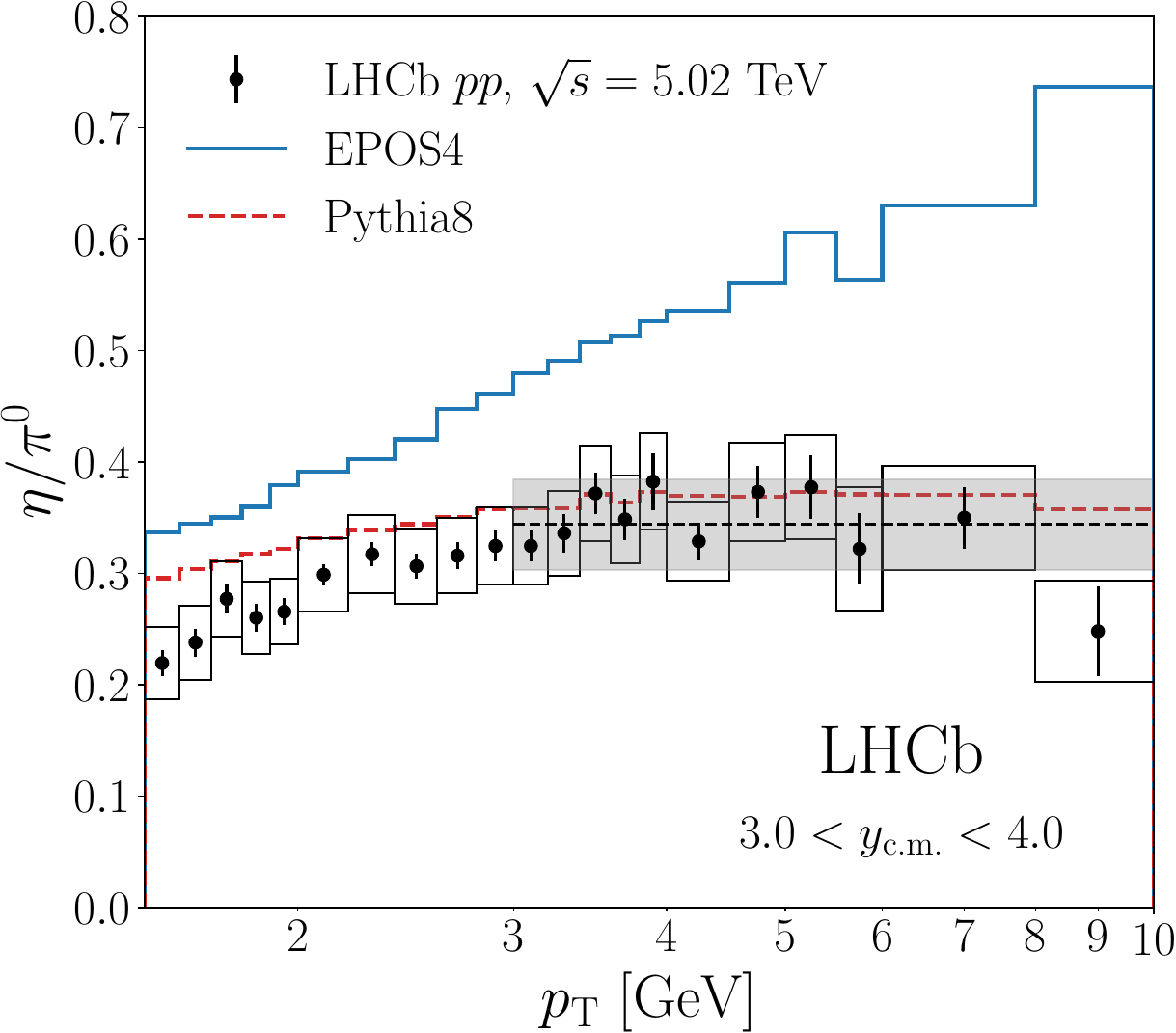}
    \includegraphics[width=0.48\textwidth]{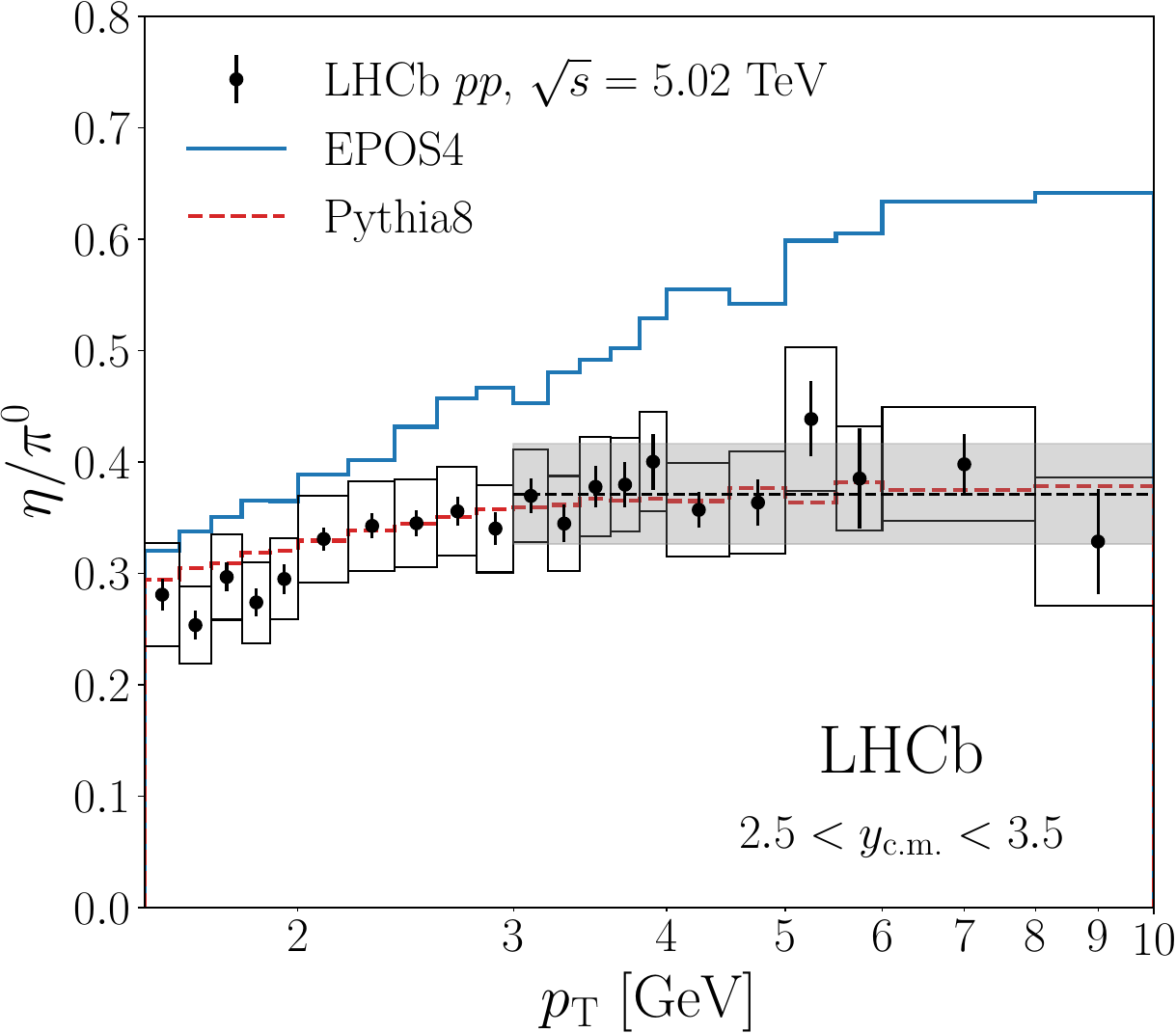}

    \includegraphics[width=0.48\textwidth]{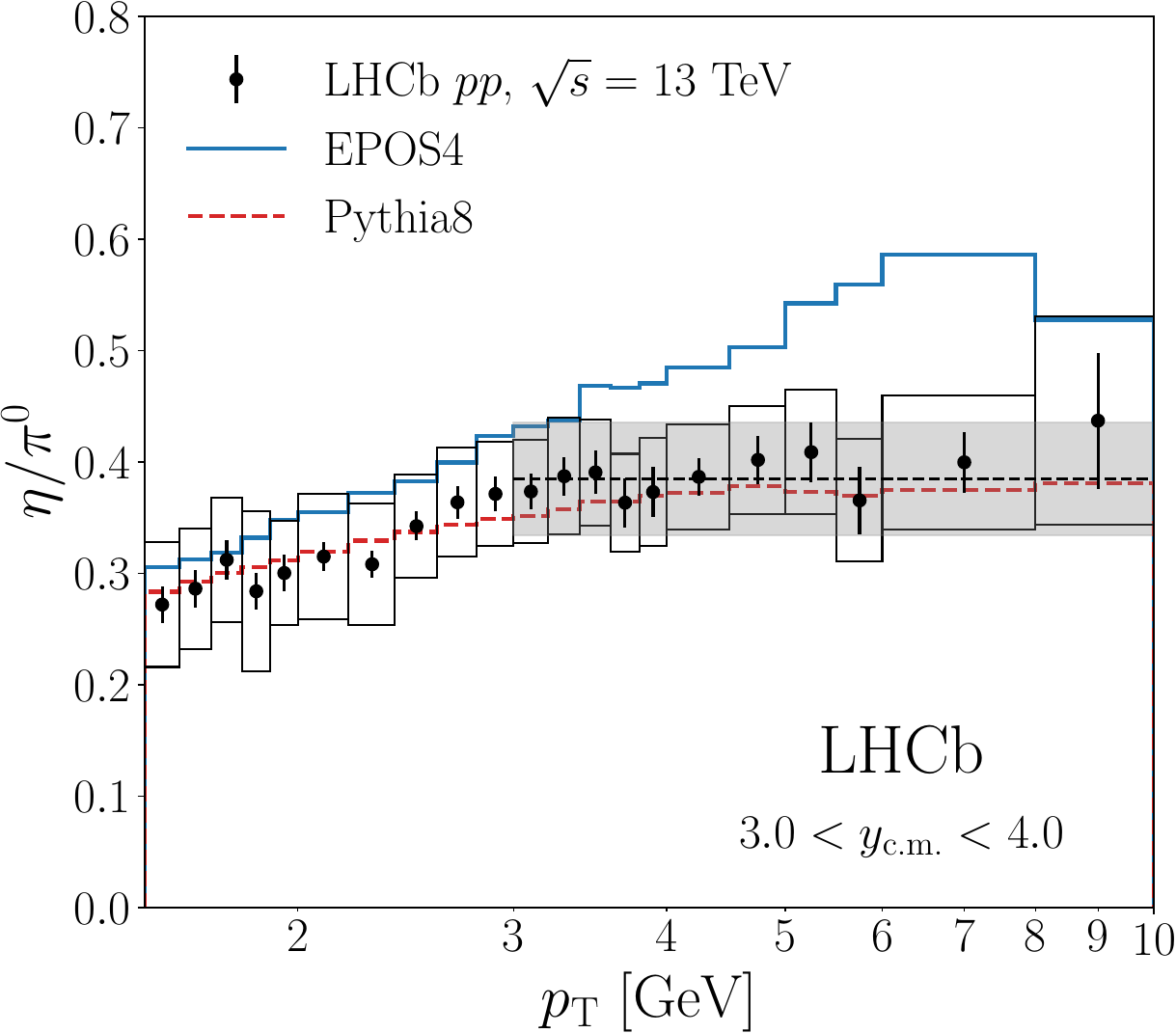}
    \includegraphics[width=0.48\textwidth]{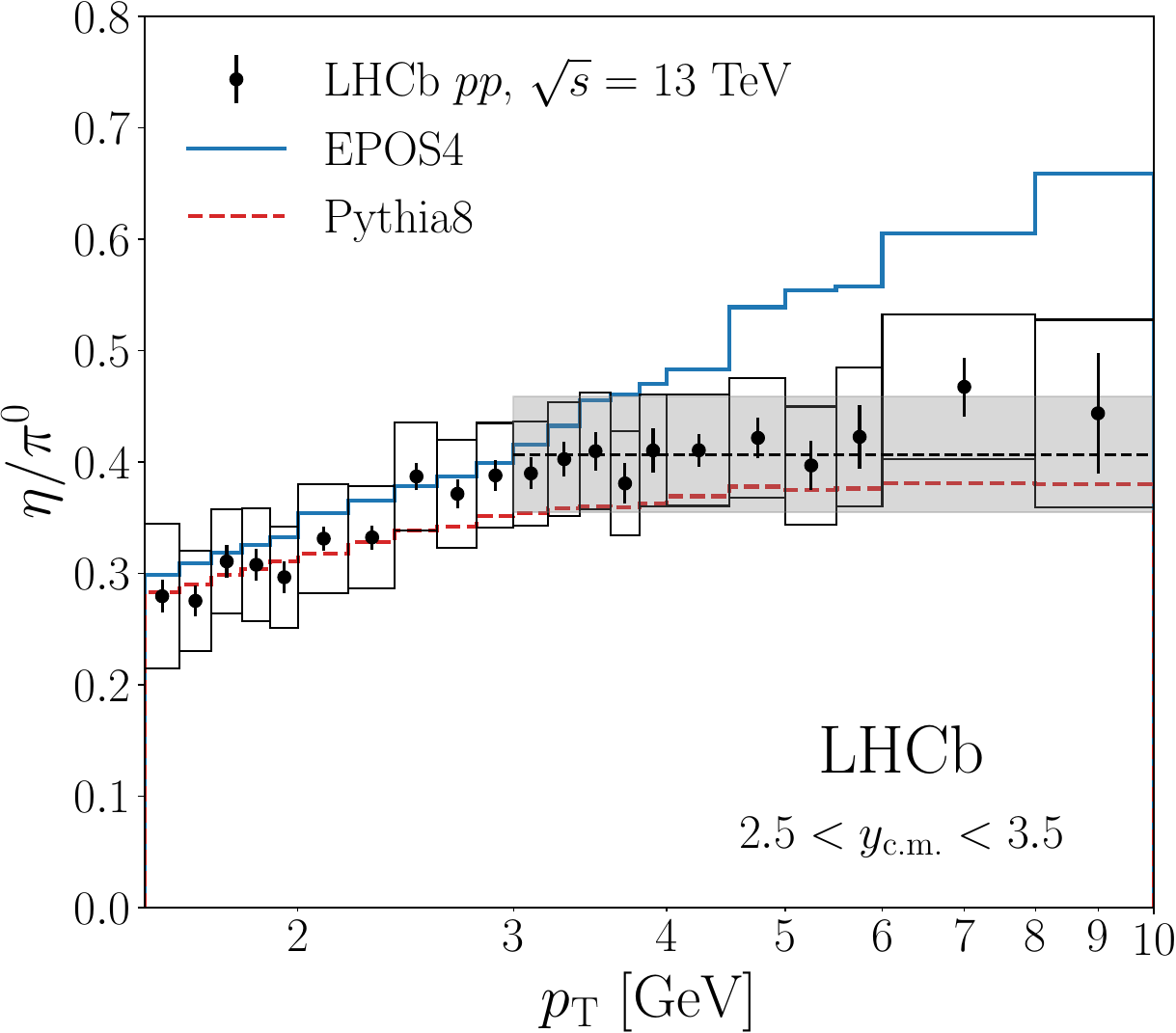}

    \includegraphics[width=0.48\textwidth]{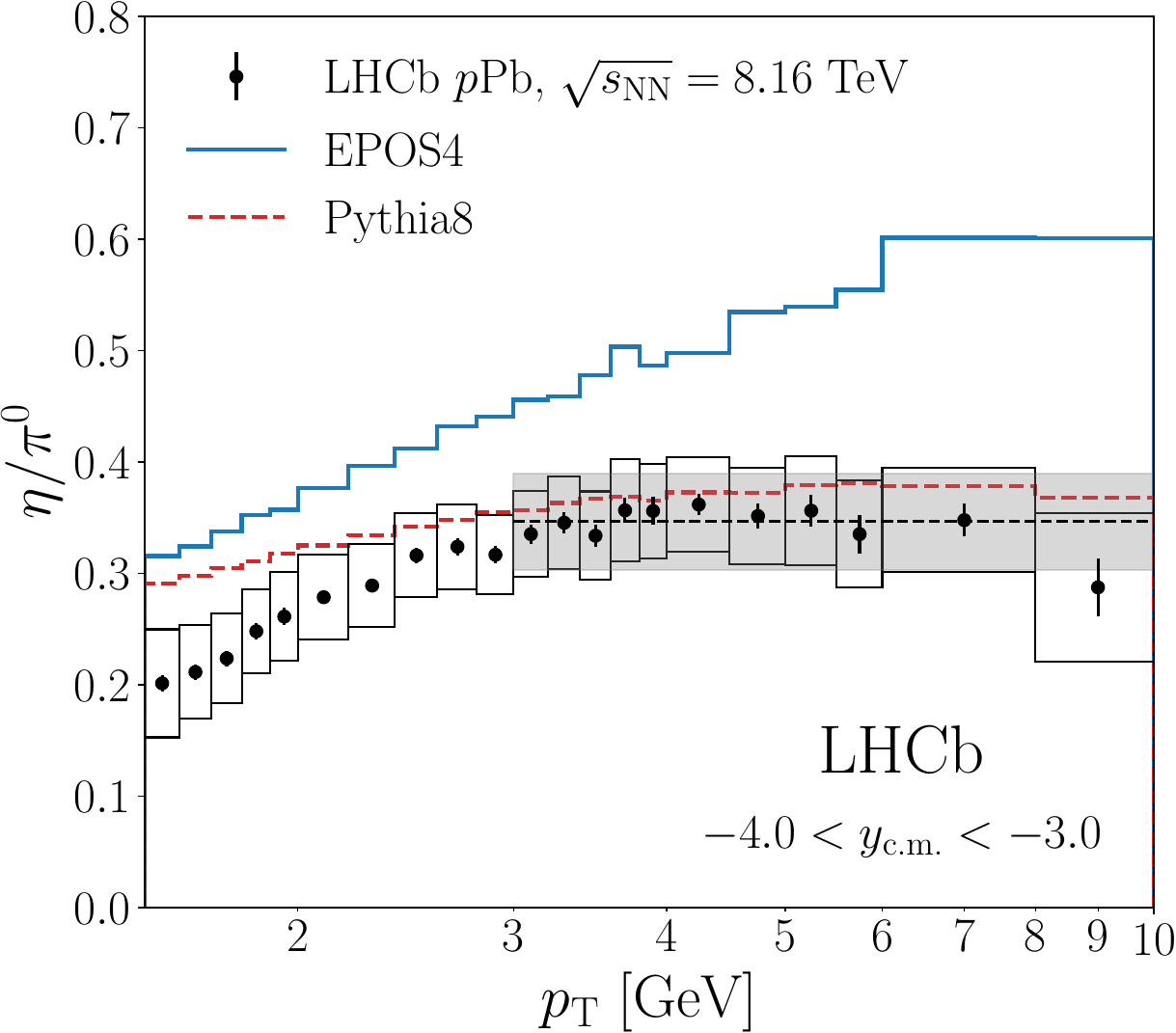}
    \includegraphics[width=0.48\textwidth]{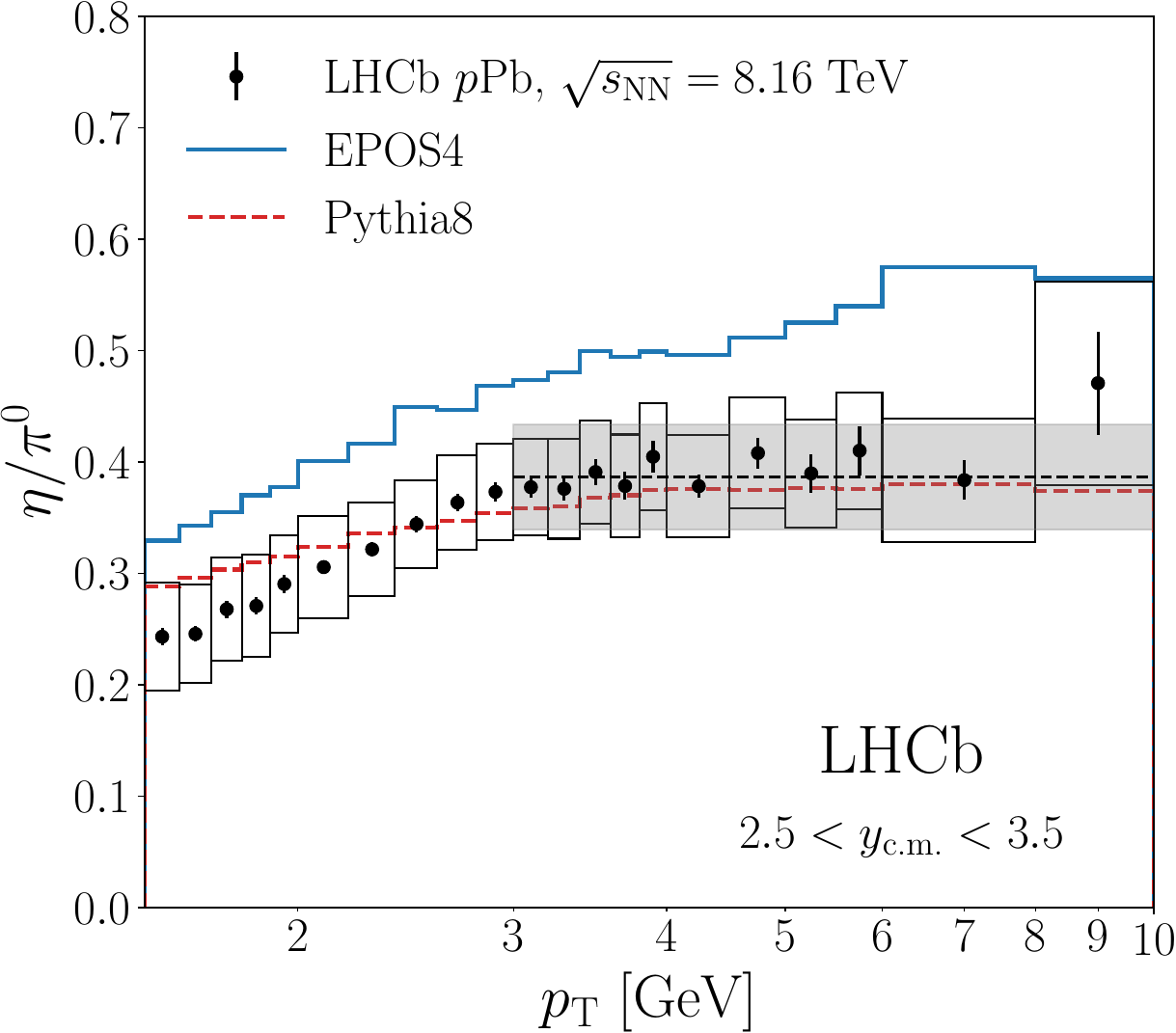}
    \caption{Measured $\eta/\pi^0$ cross section ratios in the (left) backward
    and (right) forward regions. Results are shown for $pp$ collisions at (top)
    $\sqrt{s}=5.02$ and (middle) $13~{\rm TeV}$ and (bottom) for $p{\rm Pb}$
    collisions at $\sqrt{s_{\rm NN}}=8.16~{\rm TeV}$. Results are compared to
    predictions from EPOS4 and \pythia{}8. The statistical uncertainties are
    shown by error bars, while systematic uncertainties are shown by boxes. The
    systematic uncertainties are approximately fully correlated between $p_{\rm
    T}$ regions. The black dashed line and the gray shaded region show the
    central value of $C^{\eta/\pi^0}$ and its uncertainty, respectively, where
    $C^{\eta/\pi^0}$ is calculated for \mbox{$p_{\rm T}>3\gev{}$}.}
    \label{fig:etapi0}
\end{figure}

The measured $\eta$ differential cross sections are combined with $\pi^0$
differential cross sections from Ref.~\cite{LHCb-PAPER-2021-053} to calculate
$\eta/\pi^0$ cross section ratios. These ratios are presented in
Fig.~\ref{fig:etapi0} and tabulated in Tables~\ref{tab:etapi0_forward} and
\ref{tab:etapi0_backward}. The $\eta$ measurement was performed using only
photons reconstructed as ECAL clusters, while the $\pi^0$ measurements were
performed using combinations of ECAL photons and converted photons. Furthermore,
the two measurements use different selection criteria for ECAL photons. As a
result, the systematic uncertainties between the two measurements are treated as
uncorrelated except for the luminosity uncertainty, which cancels in the cross
section ratio. The measured $\eta/\pi^0$ ratios are compared to predictions from
\pythia{}8 and EPOS4. \pythia{}8 generally describes the data well, while EPOS4
generally overestimates the $\eta/\pi^0$ ratio, especially at high $p_{\rm T}$. 

The $\eta/\pi^0$ ratio tends to plateau at $p_{\rm T}$ above a few $\gev{}$, and
the ratio is often characterized by its plateau height $C^{\eta/\pi^0}$, which
is calculated here as the average $\eta/\pi^0$ ratio for \mbox{$p_{\rm
T}>3\gev{}$}. This plateau height is illustrated in Fig.~\ref{fig:etapi0} and
tabulated in Table~\ref{tab:cetapi0}. Previous studies of the $\eta/\pi^0$ ratio
find $C^{\eta/\pi^0}$ values of $0.45-0.5$, regardless of the species of the
colliding nuclei and the center-of-mass energy of the
collision~\cite{Ren:2021pzi}. This robustness to changes in experimental
conditions illustrates the universality of fragmentation functions in hadron
collisions. The values of $C^{\eta/\pi^0}$ from this study are lower than the
universal average, with $C^{\eta/\pi^0}$ decreasing as absolute rapidity
increases and as $\sqrt{s_{\rm (NN)}}$ decreases. The ratio of the $\eta$ meson
fragmentation function to that of the $\pi^0$ meson differs with the fragmenting
parton species, so the variation seen in this measurement can be explained by
changes in the flavor of produced partons due to the kinematic dependence of
parton distribution functions~\cite{Aidala:2010bn}. Furthermore, the ratio of
fragmentation functions varies with $z$, the momentum fraction of the parton
carried by the fragmentation product. The data presented here occupy an extreme
kinematic regime and provide access to previously unexplored combinations of $z$
and initial-state parton densities. As a result, these data will provide new
constraints in future global analyses of the $\eta$ meson fragmentation
function.

\begin{table}
\centering
\caption{ Measured $C^{\eta/\pi^0}$ for each data set and rapidity region. The
first uncertainty is statistical and the second is systematic. The systematic
uncertainties are approximately 90\% positively correlated between data sets.}
\begin{tabular}{c | c c}
& $-4.0<y_{\rm c.m.}<-3.0$ & $2.5<y_{\rm c.m.}<3.5$\\
\hline
$5.02$ TeV $pp$ & $0.344\pm0.006\pm0.040$ & $0.371\pm0.006\pm0.045$\\ 
$13$ TeV $pp$ & $0.385\pm0.006\pm0.050$ & $0.407\pm0.006\pm0.052$\\ 
$8.16$ TeV $p{\rm Pb}$ & $0.346\pm0.004\pm0.043$ & $0.386\pm0.004\pm0.047$\\ 
\end{tabular}
\label{tab:cetapi0}
\end{table}

\section{Summary}

This article reports the $\eta$ and $\eta'$ meson differential cross sections at
forward and backward rapidity in $pp$ collisions at $\sqrt{s}=5.02$ and $13~{\rm
TeV}$, as well as in $p{\rm Pb}$ collisions at $\sqrt{s_{\rm NN}}=8.16~{\rm
TeV}$ collected by the LHCb experiment. This is the first study of $\eta$ meson
production at forward and backward rapidity at the LHC and the first study of
$\eta'$ meson production in high-energy proton-ion collisions. The measured
differential cross sections are compared to predictions from \pythia{}8 and
EPOS4. Neither event generator successfully describes the measurements for every
data set and rapidity region. This disagreement reflects a lack of previously
available light-hadron production data at LHC energies and forward rapidity. 

The $\eta$ meson differential cross sections are used to calculate the
$\eta/\pi^0$ cross section ratio. The measured $\eta/\pi^0$ ratios show evidence
of deviation from the universal behavior observed at central rapidity. This
deviation indicates that these data are sensitive to the $\eta$ fragmentation
functions in a complementary kinematic regime to previous studies of $\eta$
production in hadron collisions. The $\eta$ and $\eta'$ nuclear modification
factors are also reported. The measured nuclear modification factors of the
$\pi^0$, $\eta$, and $\eta'$ mesons all agree at both forward and backward
rapidity for $p_{\rm T}>3\gev{}$. These data provide limits on the mass
dependence of nuclear effects such as radial flow in $p{\rm Pb}$ collisions.
Consequently, these data will aid in the interpretation of baryon and
strangeness enhancement studies in small collision systems.

\clearpage

\section*{Acknowledgements}
%
%
\noindent We express our gratitude to our colleagues in the CERN
accelerator departments for the excellent performance of the LHC. We
thank the technical and administrative staff at the LHCb
institutes.
We acknowledge support from CERN and from the national agencies:
CAPES, CNPq, FAPERJ and FINEP (Brazil); 
MOST and NSFC (China); 
CNRS/IN2P3 (France); 
BMBF, DFG and MPG (Germany); 
INFN (Italy); 
NWO (Netherlands); 
MNiSW and NCN (Poland); 
MCID/IFA (Romania); 
MICINN (Spain); 
SNSF and SER (Switzerland); 
NASU (Ukraine); 
STFC (United Kingdom); 
DOE NP and NSF (USA).
We acknowledge the computing resources that are provided by CERN, IN2P3
(France), KIT and DESY (Germany), INFN (Italy), SURF (Netherlands),
PIC (Spain), GridPP (United Kingdom), 
CSCS (Switzerland), IFIN-HH (Romania), CBPF (Brazil),
and Polish WLCG (Poland).
We are indebted to the communities behind the multiple open-source
software packages on which we depend.
Individual groups or members have received support from
ARC and ARDC (Australia);
Key Research Program of Frontier Sciences of CAS, CAS PIFI, CAS CCEPP, 
Fundamental Research Funds for the Central Universities, 
and Sci. \& Tech. Program of Guangzhou (China);
Minciencias (Colombia);
EPLANET, Marie Sk\l{}odowska-Curie Actions, ERC and NextGenerationEU (European Union);
A*MIDEX, ANR, IPhU and Labex P2IO, and R\'{e}gion Auvergne-Rh\^{o}ne-Alpes (France);
AvH Foundation (Germany);
ICSC (Italy); 
GVA, XuntaGal, GENCAT, Inditex, InTalent and Prog.~Atracci\'on Talento, CM (Spain);
SRC (Sweden);
the Leverhulme Trust, the Royal Society
 and UKRI (United Kingdom).

\clearpage


\section*{Appendix}

\begin{table}[h]
    \centering

    \caption{Measured $d\sigma/dp_{\rm T}$ of $\eta$ production in $pp$
    collisions at \mbox{$\sqrt{s}=5.02~{\rm TeV}$}. Results and uncertainties
    are given in ${\rm mb/GeV}$. The first uncertainty is statistical and the
    second is systematic. The statistical uncertainty is uncorrelated between
    $p_{\rm T}$ intervals, while the systematic uncertainty is approximately
    fully correlated.}

    \begin{tabular}{c | c c}
    $p_{\rm T}$ (GeV) & $2.5<y_{\rm c.m.}<3.5$ & $3.0<y_{\rm c.m.}<4.0$\\
    \hline
$1.5-1.6$ &   $2.34\pm0.11\pm0.32$                 & $1.65\pm0.09\pm0.21$       \\
$1.6-1.7$ &   $1.68\pm0.08\pm0.17$                 & $1.41\pm0.07\pm0.16$       \\
$1.7-1.8$ &   $1.48\pm0.06\pm0.17$                 & $1.23\pm0.05\pm0.13$       \\
$1.8-1.9$ &   $1.09\pm0.05\pm0.13$                 & $0.90\pm0.04\pm0.10$      \\
$1.9-2.0$ &   $0.928\pm0.038\pm0.090$              & $0.742\pm0.031\pm0.069$        \\
$2.0-2.2$ &   $0.731\pm0.021\pm0.071$              & $0.572\pm0.017\pm0.053$        \\
$2.2-2.4$ &   $0.505\pm0.015\pm0.049$              & $0.400\pm0.012\pm0.037$      \\
$2.4-2.6$ &   $0.348\pm0.011\pm0.033$              & $0.265\pm0.008\pm0.025$        \\
$2.6-2.8$ &   $0.254\pm0.008\pm0.023$              & $0.186\pm0.006\pm0.017$        \\
$2.8-3.0$ &   $0.172\pm0.007\pm0.017$              & $0.136\pm0.005\pm0.012$        \\
$3.0-3.2$ &   $0.139\pm0.005\pm0.013$              & $(9.87\pm0.36\pm0.89)\times10^{-2}$     \\
$3.2-3.4$ &   $(9.42\pm0.39\pm0.99)\times10^{-2}$  & $(7.38\pm0.32\pm0.69)\times10^{-2}$     \\
$3.4-3.6$ &   $(7.33\pm0.29\pm0.67)\times10^{-2}$  & $(6.00\pm0.24\pm0.54)\times10^{-2}$       \\
$3.6-3.8$ &   $(6.06\pm0.25\pm0.55)\times10^{-2}$  & $(4.47\pm0.19\pm0.41)\times10^{-2}$     \\
$3.8-4.0$ &   $(4.67\pm0.24\pm0.43)\times10^{-2}$  & $(3.53\pm0.19\pm0.32)\times10^{-2}$     \\
$4.0-4.5$ &   $(2.84\pm0.11\pm0.26)\times10^{-2}$  & $(2.06\pm0.08\pm0.19)\times10^{-2}$     \\
$4.5-5.0$ &   $(1.67\pm0.08\pm0.16)\times10^{-2}$  & $(1.23\pm0.06\pm0.12)\times10^{-2}$     \\
$5.0-5.5$ &   $(1.05\pm0.06\pm0.10)\times10^{-2}$  & $(7.80\pm0.43\pm0.74)\times10^{-3}$       \\
$5.5-6.0$ &   $(5.97\pm0.61\pm0.58)\times10^{-3}$  & $(3.84\pm0.29\pm0.61)\times10^{-3}$      \\
$6.0-8.0$ &   $(2.49\pm0.13\pm0.24)\times10^{-3}$  & $(1.75\pm0.11\pm0.17)\times10^{-3}$      \\
$8.0-10.0$ &  $(5.67\pm0.67\pm0.74)\times10^{-4}$  & $(3.72\pm0.49\pm0.42)\times10^{-4}$     \\
    \end{tabular}
    
    \label{tab:eta_xsec_pp2510}
\end{table}

\begin{table}[h]
    \centering

    \caption{Measured $d\sigma/dp_{\rm T}$ of $\eta$ production in $pp$
    collisions at \mbox{$\sqrt{s}=13~{\rm TeV}$}. Results and uncertainties are
    given in ${\rm mb/GeV}$. The first uncertainty is statistical and the second
    is systematic. The statistical uncertainty is uncorrelated between $p_{\rm
    T}$ intervals, while the systematic uncertainty is approximately fully
    correlated.}

    \begin{tabular}{c | c c}
    $p_{\rm T}$ (GeV) & $2.5<y_{\rm c.m.}<3.5$ & $3.0<y_{\rm c.m.}<4.0$\\
    \hline
$1.5-1.6$  & $3.76\pm0.19\pm0.76$ &  $3.17\pm0.19\pm0.68$\\
$1.6-1.7$  & $2.95\pm0.14\pm0.42$ &  $2.66\pm0.15\pm0.46$\\
$1.7-1.8$  & $2.56\pm0.12\pm0.38$ &  $2.18\pm0.12\pm0.39$\\
$1.8-1.9$  & $2.05\pm0.09\pm0.29$ &  $1.61\pm0.09\pm0.34$\\
$1.9-2.0$  & $1.62\pm0.07\pm0.22$  &  $1.40\pm0.07\pm0.20$\\
$2.0-2.2$  & $1.30\pm0.04\pm0.17$  &  $1.03\pm0.04\pm0.17$\\
$2.2-2.4$  & $0.88\pm0.03\pm0.11$  &  $0.70\pm0.03\pm0.11$\\
$2.4-2.6$  & $0.719\pm0.021\pm0.083$  &  $0.531\pm0.019\pm0.069$\\
$2.6-2.8$  & $0.508\pm0.016\pm0.059$  &  $0.397\pm0.015\pm0.050$\\
$2.8-3.0$  & $0.387\pm0.012\pm0.043$  &  $0.297\pm0.012\pm0.034$\\
$3.0-3.2$  & $0.294\pm0.009\pm0.032$  &  $0.221\pm0.008\pm0.025$\\
$3.2-3.4$  & $0.230\pm0.008\pm0.025$  &  $0.175\pm0.007\pm0.021$\\
$3.4-3.6$  & $0.171\pm0.006\pm0.019$  &  $0.128\pm0.006\pm0.014$\\
$3.6-3.8$  & $0.127\pm0.005\pm0.013$  &  $(9.55\pm0.51\pm0.93)\times10^{-2}$\\
$3.8-4.0$  & $0.103\pm0.004\pm0.010$  &  $(7.36\pm0.38\pm0.83)\times10^{-2}$\\
$4.0-4.5$  & $(7.04\pm0.20\pm0.75)\times10^{-2}$  &  $(5.09\pm0.18\pm0.53)\times10^{-2}$\\
$4.5-5.0$  & $(4.32\pm0.15\pm0.43)\times10^{-2}$  &  $(3.01\pm0.14\pm0.32)\times10^{-2}$\\
$5.0-5.5$  & $(2.60\pm0.11\pm0.27)\times10^{-2}$  &  $(2.01\pm0.11\pm0.23)\times10^{-2}$\\
$5.5-6.0$  & $(1.63\pm0.08\pm0.18)\times10^{-2}$  &  $(1.05\pm0.07\pm0.14)\times10^{-2}$\\
$6.0-8.0$  & $(6.86\pm0.31\pm0.74)\times10^{-3}$  &  $(4.72\pm0.26\pm0.65)\times10^{-3}$\\
$8.0-10.0$ & $(2.01\pm0.21\pm0.29)\times10^{-3}$  &  $(1.50\pm0.17\pm0.27)\times10^{-3}$\\
    \end{tabular}
    
    \label{tab:eta_xsec_pp6500}
\end{table}

\begin{table}[h]
    \centering

    \caption{Measured $d\sigma/dp_{\rm T}$ of $\eta$ production in $p{\rm Pb}$
    collisions at \mbox{$\sqrt{s_{\rm NN}}=8.16~{\rm TeV}$}. Results and
    uncertainties are given in ${\rm mb/GeV}$. The first uncertainty is
    statistical and the second is systematic. The statistical uncertainty is
    uncorrelated between $p_{\rm T}$ intervals, while the systematic uncertainty
    is approximately fully correlated.}

    \begin{tabular}{c | c c}
    $p_{\rm T}$ (GeV) & $2.5<y_{\rm c.m.}<3.5$ & $-4.0<y_{\rm c.m.}<-3.0$\\
    \hline
$1.5-1.6$ & $342\pm10\pm61$ & $387\pm12\pm85$\\
$1.6-1.7$ & $274\pm7\pm44$ & $320\pm9\pm59$\\
$1.7-1.8$ & $237\pm6\pm38$ & $271\pm8\pm45$\\
$1.8-1.9$ & $193\pm5\pm31$ & $237\pm7\pm32$\\
$1.9-2.0$ & $171\pm4\pm24$ & $201\pm5\pm28$\\
$2.0-2.2$ & $130\pm2\pm18$ & $162\pm3\pm20$\\
$2.2-2.4$ & $97\pm2\pm11$ & $113\pm2\pm13$\\
$2.4-2.6$ & $72.4\pm1.2\pm7.8$ & $87.8\pm1.6\pm9.1$\\
$2.6-2.8$ & $54.1\pm0.9\pm5.5$ & $63.0\pm1.2\pm6.4$\\
$2.8-3.0$ & $40.4\pm0.7\pm4.1$ & $44.5\pm0.9\pm4.3$\\
$3.0-3.2$ & $30.7\pm0.6\pm3.1$ & $34.9\pm0.7\pm3.5$\\
$3.2-3.4$ & $23.5\pm0.4\pm2.4$ & $26.5\pm0.6\pm2.7$\\
$3.4-3.6$ & $18.2\pm0.4\pm1.8$ & $18.9\pm0.4\pm1.9$\\
$3.6-3.8$ & $14.2\pm0.3\pm1.5$ & $15.2\pm0.4\pm1.6$\\
$3.8-4.0$ & $11.2\pm0.3\pm1.2$ & $11.9\pm0.3\pm1.2$\\
$4.0-4.5$ & $7.39\pm0.13\pm0.77$ & $7.77\pm0.15\pm0.77$\\
$4.5-5.0$ & $4.39\pm0.09\pm0.45$ & $4.31\pm0.10\pm0.41$\\
$5.0-5.5$ & $2.74\pm0.07\pm0.27$ & $2.61\pm0.07\pm0.26$\\
$5.5-6.0$ & $1.66\pm0.06\pm0.17$ & $1.61\pm0.05\pm0.16$\\
$6.0-8.0$ & $0.722\pm0.021\pm0.077$ & $0.655\pm0.018\pm0.060$\\
$8.0-10.0$ &  $0.237\pm0.016\pm0.039$ & $0.159\pm0.010\pm0.028$\\
    \end{tabular}
    
    \label{tab:eta_xsec_pPb}
\end{table}

\begin{table}[h]
    \centering

    \caption{Measured $d\sigma/dp_{\rm T}$ of $\eta'$ production in $pp$
    collisions at \mbox{$\sqrt{s}=5.02~{\rm TeV}$}. Results and uncertainties
    are given in ${\rm mb/GeV}$. The first uncertainty is statistical and the
    second is systematic. The statistical uncertainty is uncorrelated between
    $p_{\rm T}$ bins, while the systematic uncertainty is approximately fully
    correlated.}    

    \begin{tabular}{c | c c}
    $p_{\rm T}$ (GeV) & $2.5<y_{\rm c.m.}<3.5$ & $3.0<y_{\rm c.m.}<4.0$\\
    \hline
$3-4$ &   $(2.64\pm0.50\pm0.84)\times10^{-2}$ & $(2.04\pm0.32\pm0.74)\times10^{-2}$\\  
$4-5$ &   $(9.0\pm1.3\pm1.9)\times10^{-3}$  & $(6.7\pm0.9\pm2.2)\times10^{-3}$ \\
$5-7$ &   $(1.95\pm0.10\pm0.45)\times10^{-3}$ & $(1.67\pm0.24\pm0.51)\times10^{-3}$\\  
$7-10$ &  $(3.75\pm0.96\pm0.85)\times10^{-4}$ & $(2.89\pm0.76\pm0.74)\times10^{-4}$\\
    \end{tabular}
    
    \label{tab:etap_xsec_pp2510}
\end{table}

\begin{table}[h]
    \centering

    \caption{Measured $d\sigma/dp_{\rm T}$ of $\eta'$ production in $pp$
    collisions at \mbox{$\sqrt{s}=13~{\rm TeV}$}. Results and uncertainties are
    given in ${\rm mb/GeV}$. The first uncertainty is statistical and the second
    is systematic. The statistical uncertainty is uncorrelated between $p_{\rm
    T}$ bins, while the systematic uncertainty is approximately fully
    correlated.}   

    \begin{tabular}{c | c c}
    $p_{\rm T}$ (GeV) & $2.5<y_{\rm c.m.}<3.5$ & $3.0<y_{\rm c.m.}<4.0$\\
    \hline
$3-4$ &  $(5.7\pm0.9\pm1.5)\times10^{-2}$    & $(4.0\pm0.6\pm1.1)\times10^{-2}$ \\
$4-5$ &  $(2.05\pm0.23\pm0.58)\times10^{-2}$ & $(1.51\pm0.15\pm0.53)\times10^{-2}$ \\
$5-7$ &  $(6.5\pm0.5\pm1.3)\times10^{-3}$    & $(4.33\pm0.39\pm0.92)\times10^{-3}$ \\
$7-10$ & $(1.61\pm0.16\pm0.41)\times10^{-3}$ & $(9.1\pm1.2\pm2.1)\times10^{-4}$ \\
    \end{tabular}
     
    \label{tab:etap_xsec_pp6500}
\end{table}

\begin{table}[h]
    \centering

    \caption{Measured $d\sigma/dp_{\rm T}$ of $\eta'$ production in $p{\rm Pb}$
    collisions at \mbox{$\sqrt{s_{\rm NN}}=8.16~{\rm TeV}$}. Results and
    uncertainties are given in ${\rm mb/GeV}$. The first uncertainty is
    statistical and the second is systematic. The statistical uncertainty is
    uncorrelated between $p_{\rm T}$ bins, while the systematic uncertainty is
    approximately fully correlated.}

    \begin{tabular}{c | c c}
    $p_{\rm T}$ (GeV) & $2.5<y_{\rm c.m.}<3.5$ & $-4.0<y_{\rm c.m.}<-3.0$\\
    \hline
$3-4$ &   $5.8\pm0.5\pm2.1$ & $6.0\pm1.0\pm1.6$\\
$4-5$ &   $1.82\pm0.12\pm0.56$ & $2.14\pm0.2\pm0.49$\\
$5-7$ &   $0.61\pm0.03\pm0.15$ & $0.68\pm0.04\pm0.17$\\
$7-10$ &  $0.128\pm0.009\pm0.027$ & $0.111\pm0.008\pm0.018$\\
    \end{tabular}
    
    \label{tab:etap_xsec_pPb}
\end{table}

\begin{table}[h]
    \centering

    \caption{Measured $\eta/\pi^0$ cross section ratio for $2.5<y_{\rm
    c.m.}<3.5$. The first uncertainty is statistical and the second is
    systematic. The statistical uncertainty is uncorrelated between $p_{\rm T}$
    bins, while the systematic uncertainty is approximately fully correlated.}

    \begin{tabular}{c | c c c}
    $p_{\rm T}$ (GeV) & $5.02$ TeV $pp$ & $13$ TeV $pp$ & $8.16$ TeV $p{\rm Pb}$\\
    \hline
$1.5-1.6$ & $0.281\pm0.014\pm0.047$ & $0.279\pm0.015\pm0.060$ & $0.246\pm0.008\pm0.049$\\ 
$1.6-1.7$ & $0.253\pm0.013\pm0.035$ & $0.275\pm0.014\pm0.045$ & $0.248\pm0.007\pm0.045$\\ 
$1.7-1.8$ & $0.297\pm0.013\pm0.038$ & $0.311\pm0.015\pm0.050$ & $0.270\pm0.008\pm0.047$\\ 
$1.8-1.9$ & $0.274\pm0.013\pm0.036$ & $0.308\pm0.014\pm0.047$ & $0.273\pm0.008\pm0.046$\\ 
$1.9-2.0$ & $0.295\pm0.013\pm0.036$ & $0.297\pm0.014\pm0.045$ & $0.293\pm0.008\pm0.044$\\ 
$2.0-2.2$ & $0.331\pm0.011\pm0.039$ & $0.331\pm0.011\pm0.048$ & $0.309\pm0.006\pm0.046$\\ 
$2.2-2.4$ & $0.343\pm0.011\pm0.040$ & $0.332\pm0.011\pm0.046$ & $0.325\pm0.007\pm0.043$\\ 
$2.4-2.6$ & $0.345\pm0.012\pm0.039$ & $0.387\pm0.012\pm0.048$ & $0.348\pm0.007\pm0.040$\\ 
$2.6-2.8$ & $0.356\pm0.013\pm0.040$ & $0.371\pm0.013\pm0.048$ & $0.367\pm0.008\pm0.043$\\ 
$2.8-3.0$ & $0.340\pm0.015\pm0.039$ & $0.388\pm0.014\pm0.048$ & $0.377\pm0.009\pm0.043$\\ 
$3.0-3.2$ & $0.370\pm0.015\pm0.042$ & $0.390\pm0.014\pm0.047$ & $0.381\pm0.010\pm0.044$\\ 
$3.2-3.4$ & $0.345\pm0.017\pm0.043$ & $0.402\pm0.016\pm0.051$ & $0.380\pm0.010\pm0.045$\\ 
$3.4-3.6$ & $0.378\pm0.018\pm0.045$ & $0.410\pm0.017\pm0.053$ & $0.395\pm0.012\pm0.047$\\ 
$3.6-3.8$ & $0.380\pm0.020\pm0.042$ & $0.381\pm0.018\pm0.046$ & $0.382\pm0.013\pm0.047$\\ 
$3.8-4.0$ & $0.400\pm0.025\pm0.045$ & $0.410\pm0.020\pm0.050$ & $0.409\pm0.015\pm0.049$\\ 
$4.0-4.5$ & $0.357\pm0.016\pm0.042$ & $0.410\pm0.015\pm0.050$ & $0.382\pm0.011\pm0.046$\\ 
$4.5-5.0$ & $0.364\pm0.021\pm0.046$ & $0.422\pm0.018\pm0.053$ & $0.412\pm0.014\pm0.050$\\ 
$5.0-5.5$ & $0.439\pm0.034\pm0.064$ & $0.397\pm0.022\pm0.053$ & $0.394\pm0.017\pm0.049$\\ 
$5.5-6.0$ & $0.385\pm0.045\pm0.047$ & $0.423\pm0.028\pm0.062$ & $0.414\pm0.022\pm0.053$\\ 
$6.0-8.0$ & $0.398\pm0.027\pm0.051$ & $0.467\pm0.026\pm0.064$ & $0.388\pm0.018\pm0.056$\\ 
$8.0-10.0$ & $0.329\pm0.047\pm0.057$ & $0.444\pm0.054\pm0.079$ & $0.475\pm0.047\pm0.092$\\ 
    \end{tabular}
    
    \label{tab:etapi0_forward}
\end{table}

\begin{table}[h]
    \centering
    
    \caption{Measured $\eta/\pi^0$ cross section ratio for $-4.0<y_{\rm
    c.m.}<-3.0$. The first uncertainty is statistical and the second is
    systematic. The statistical uncertainty is uncorrelated between $p_{\rm T}$
    bins, while the systematic uncertainty is approximately fully correlated.}

    \begin{tabular}{c | c c c}
    $p_{\rm T}$ (GeV) & $5.02$ TeV $pp$ & $13$ TeV $pp$ & $8.16$ TeV $p{\rm Pb}$\\
    \hline
$1.5-1.6$ & $0.219\pm0.012\pm0.032$ & $0.272\pm0.016\pm0.061$ & $0.197\pm0.007\pm0.047$\\ 
$1.6-1.7$ & $0.238\pm0.013\pm0.033$ & $0.286\pm0.017\pm0.053$ & $0.209\pm0.007\pm0.042$\\ 
$1.7-1.8$ & $0.277\pm0.013\pm0.034$ & $0.312\pm0.018\pm0.059$ & $0.224\pm0.007\pm0.040$\\ 
$1.8-1.9$ & $0.260\pm0.012\pm0.033$ & $0.284\pm0.016\pm0.062$ & $0.248\pm0.007\pm0.037$\\ 
$1.9-2.0$ & $0.266\pm0.012\pm0.029$ & $0.300\pm0.016\pm0.047$ & $0.256\pm0.007\pm0.039$\\ 
$2.0-2.2$ & $0.299\pm0.010\pm0.033$ & $0.315\pm0.013\pm0.055$ & $0.276\pm0.006\pm0.038$\\ 
$2.2-2.4$ & $0.317\pm0.011\pm0.035$ & $0.308\pm0.012\pm0.051$ & $0.286\pm0.006\pm0.036$\\ 
$2.4-2.6$ & $0.306\pm0.011\pm0.034$ & $0.342\pm0.013\pm0.048$ & $0.314\pm0.007\pm0.037$\\ 
$2.6-2.8$ & $0.316\pm0.012\pm0.034$ & $0.364\pm0.015\pm0.049$ & $0.323\pm0.007\pm0.038$\\ 
$2.8-3.0$ & $0.325\pm0.014\pm0.035$ & $0.371\pm0.016\pm0.047$ & $0.320\pm0.008\pm0.036$\\ 
$3.0-3.2$ & $0.325\pm0.014\pm0.035$ & $0.373\pm0.016\pm0.047$ & $0.336\pm0.009\pm0.038$\\ 
$3.2-3.4$ & $0.336\pm0.017\pm0.038$ & $0.387\pm0.017\pm0.053$ & $0.345\pm0.009\pm0.041$\\ 
$3.4-3.6$ & $0.372\pm0.019\pm0.043$ & $0.391\pm0.020\pm0.048$ & $0.337\pm0.010\pm0.040$\\ 
$3.6-3.8$ & $0.348\pm0.019\pm0.040$ & $0.363\pm0.022\pm0.042$ & $0.357\pm0.011\pm0.046$\\ 
$3.8-4.0$ & $0.382\pm0.026\pm0.043$ & $0.373\pm0.022\pm0.048$ & $0.359\pm0.012\pm0.043$\\ 
$4.0-4.5$ & $0.329\pm0.017\pm0.035$ & $0.387\pm0.017\pm0.048$ & $0.364\pm0.009\pm0.042$\\ 
$4.5-5.0$ & $0.373\pm0.023\pm0.044$ & $0.402\pm0.022\pm0.050$ & $0.352\pm0.011\pm0.044$\\ 
$5.0-5.5$ & $0.377\pm0.029\pm0.047$ & $0.409\pm0.027\pm0.056$ & $0.357\pm0.014\pm0.050$\\ 
$5.5-6.0$ & $0.322\pm0.032\pm0.056$ & $0.365\pm0.030\pm0.054$ & $0.335\pm0.017\pm0.048$\\ 
$6.0-8.0$ & $0.350\pm0.028\pm0.047$ & $0.400\pm0.027\pm0.063$ & $0.346\pm0.014\pm0.047$\\ 
$8.0-10.0$ & $0.248\pm0.040\pm0.045$ & $0.437\pm0.061\pm0.093$ & $0.281\pm0.025\pm0.065$\\ 
    \end{tabular}
    
    \label{tab:etapi0_backward}
\end{table}

\begin{table}
    \centering

    \caption{Measured $\eta$ nuclear modification factor $R_{p{\rm Pb}}$ as a
    function of $p_{\rm T}$. The first uncertainty is statistical and the second
    is systematic. The statistical uncertainty is uncorrelated between $p_{\rm
    T}$ bins, while the systematic uncertainty is approximately fully
    correlated.}    

    \begin{tabular}{c | c c}
    $p_{\rm T}$ (GeV) & $2.5<y_{\rm c.m.}<3.5$ & $-4.0<y_{\rm c.m.}<-3.0$ \\
    \hline
$1.5-1.6$ & $0.552\pm0.025\pm0.033$ & $0.807\pm0.041\pm0.067$\\ 
$1.6-1.7$ & $0.589\pm0.025\pm0.047$ & $0.790\pm0.038\pm0.052$\\ 
$1.7-1.8$ & $0.580\pm0.024\pm0.045$ & $0.791\pm0.036\pm0.046$\\ 
$1.8-1.9$ & $0.617\pm0.025\pm0.044$ & $0.939\pm0.043\pm0.054$\\ 
$1.9-2.0$ & $0.667\pm0.026\pm0.047$ & $0.943\pm0.040\pm0.058$\\ 
$2.0-2.2$ & $0.639\pm0.017\pm0.044$ & $1.005\pm0.031\pm0.049$\\ 
$2.2-2.4$ & $0.695\pm0.019\pm0.043$ & $1.015\pm0.031\pm0.048$\\ 
$2.4-2.6$ & $0.691\pm0.018\pm0.040$ & $1.117\pm0.033\pm0.042$\\ 
$2.6-2.8$ & $0.718\pm0.020\pm0.042$ & $1.106\pm0.034\pm0.043$\\ 
$2.8-3.0$ & $0.746\pm0.023\pm0.042$ & $1.056\pm0.035\pm0.039$\\ 
$3.0-3.2$ & $0.725\pm0.022\pm0.041$ & $1.125\pm0.038\pm0.040$\\ 
$3.2-3.4$ & $0.761\pm0.025\pm0.044$ & $1.110\pm0.040\pm0.048$\\ 
$3.4-3.6$ & $0.778\pm0.026\pm0.044$ & $1.028\pm0.038\pm0.048$\\ 
$3.6-3.8$ & $0.771\pm0.028\pm0.051$ & $1.112\pm0.046\pm0.059$\\ 
$3.8-4.0$ & $0.772\pm0.031\pm0.056$ & $1.111\pm0.050\pm0.051$\\ 
$4.0-4.5$ & $0.787\pm0.023\pm0.049$ & $1.143\pm0.038\pm0.047$\\ 
$4.5-5.0$ & $0.777\pm0.028\pm0.050$ & $1.066\pm0.042\pm0.036$\\ 
$5.0-5.5$ & $0.791\pm0.034\pm0.050$ & $0.994\pm0.046\pm0.047$\\ 
$5.5-6.0$ & $0.799\pm0.053\pm0.064$ & $1.210\pm0.072\pm0.095$\\ 
$6.0-8.0$ & $0.830\pm0.037\pm0.057$ & $1.084\pm0.053\pm0.064$\\ 
$8.0-10.0$ & $1.05\pm0.11\pm0.12$ & $1.01\pm0.11\pm0.15$\\ 
    \end{tabular}
    
    \label{tab:eta_rpa}
\end{table}

\begin{table}
    \centering

    \caption{Measured $\eta'$ nuclear modification factor $R_{p{\rm Pb}}$ as a
    function of $p_{\rm T}$. The first uncertainty is statistical and the second
    is systematic. The statistical uncertainty is uncorrelated between $p_{\rm
    T}$ bins, while the systematic uncertainty is approximately fully
    correlated.} 

    \begin{tabular}{c | c c}
    $p_{\rm T}$ (GeV) & $2.5<y_{\rm c.m.}<3.5$ & $-4.0<y_{\rm c.m.}<-3.0$ \\
    \hline
$3.0-4.0$ & $0.71\pm0.10\pm0.08$ & $1.00\pm0.19\pm0.18$\\ 
$4.0-5.0$ & $0.639\pm0.073\pm0.079$ & $1.01\pm0.13\pm0.18$\\ 
$5.0-7.0$ & $0.814\pm0.055\pm0.093$ & $1.21\pm0.12\pm0.15$\\ 
$7.0-10.0$ & $0.78\pm0.12\pm0.15$ & $1.02\pm0.17\pm0.30$\\ 
    \end{tabular}
    
    \label{tab:etap_rpa}
\end{table}

\clearpage

\addcontentsline{toc}{section}{References}
\bibliographystyle{LHCb}
\bibliography{main,standard,LHCb-PAPER,LHCb-CONF,LHCb-DP,LHCb-TDR}

\clearpage
 
\centerline
{\large\bf LHCb collaboration}
\begin
{flushleft}
\small
R.~Aaij$^{35}$\lhcborcid{0000-0003-0533-1952},
A.S.W.~Abdelmotteleb$^{54}$\lhcborcid{0000-0001-7905-0542},
C.~Abellan~Beteta$^{48}$,
F.~Abudin{\'e}n$^{54}$\lhcborcid{0000-0002-6737-3528},
T.~Ackernley$^{58}$\lhcborcid{0000-0002-5951-3498},
B.~Adeva$^{44}$\lhcborcid{0000-0001-9756-3712},
M.~Adinolfi$^{52}$\lhcborcid{0000-0002-1326-1264},
P.~Adlarson$^{78}$\lhcborcid{0000-0001-6280-3851},
C.~Agapopoulou$^{46}$\lhcborcid{0000-0002-2368-0147},
C.A.~Aidala$^{79}$\lhcborcid{0000-0001-9540-4988}
% ...

{\footnotesize \it

% ...
$^{35}$Nikhef National Institute for Subatomic Physics, Amsterdam, Netherlands\\
% ...
$^{44}$Instituto Galego de F{\'\i}sica de Altas Enerx{\'\i}as (IGFAE), Universidade de Santiago de Compostela, Santiago de Compostela, Spain\\
% ...
$^{46}$European Organization for Nuclear Research (CERN), Geneva, Switzerland\\
% ...
$^{48}$Physik-Institut, Universit{\"a}t Z{\"u}rich, Z{\"u}rich, Switzerland\\
% ...
$^{52}$H.H. Wills Physics Laboratory, University of Bristol, Bristol, United Kingdom\\
% ...
$^{54}$Department of Physics, University of Warwick, Coventry, United Kingdom\\
% ...
$^{57}$School of Physics and Astronomy, University of Glasgow, Glasgow, United Kingdom\\
$^{58}$Oliver Lodge Laboratory, University of Liverpool, Liverpool, United Kingdom\\
% ...
$^{66}$Syracuse University, Syracuse, NY, United States\\
% ...
$^{78}$Department of Physics and Astronomy, Uppsala University, Uppsala, Sweden, associated to $^{57}$\\
$^{79}$University of Michigan, Ann Arbor, MI, United States, associated to $^{66}$\\
% ...
}
\end{flushleft}

\end{document}